\documentclass[a4paper,11pt]{article}
\usepackage[english]{babel}
\usepackage[utf8]{inputenc}
\pdfoutput=1
\usepackage{braket}
\usepackage{physics}
\usepackage{color}
\usepackage{amssymb}
\usepackage{amsmath}
\usepackage{graphicx}
\usepackage{epstopdf}
\usepackage{subcaption}
\usepackage{comment}
\usepackage{mathtools}
\usepackage{ragged2e}
\usepackage{hyperref}
\hypersetup{
	colorlinks=true,
	linkcolor=blue,
	filecolor=cyan, 
	urlcolor=blue,
	citecolor=red,
} 
\usepackage[titletoc,toc,title]{appendix}
\numberwithin{equation}{section}
\usepackage{cleveref}
\usepackage[margin=2.5cm]{geometry}
\usepackage{cite}
\usepackage{epsfig}
\usepackage{float}
\usepackage{cancel}
\usepackage{amsfonts}
\usepackage{enumitem}
\usepackage{caption}
\usepackage{authblk}

\makeatletter
\newcommand{\vast}{\bBigg@{3}}
\newcommand{\Vast}{\bBigg@{5}}
\makeatother

\newcommand{\TTbar}{\text{T}\bar{\text{T}}}
\newcommand{\ttbar}{T\bar{{T}}}
\newcommand{\zbar}{\raisebox{0.2ex}{--}\kern-0.6em Z}

\begin{document}
	
	\title{Reflected entropy and timelike entanglement in $\TTbar$ deformed CFT$_2$s}

	\author[1]{Debarshi Basu\thanks{\noindent E-mail:~ {\tt \href{mailto:debarshi@iitk.ac.in} {debarshi@iitk.ac.in}}}}

	\author[1]{Vinayak Raj\thanks{\noindent E-mail:~ {\tt \href{mailto:vraj@iitk.ac.in} {vraj@iitk.ac.in}}}}

	\affil[1]{
		Department of Physics\\
		
		Indian Institute of Technology\\ 
		
		Kanpur 208 016, India
	}
	
	\date{}
	
	\maketitle
	
	\thispagestyle{empty}
	
	\begin{abstract}

		We develop a covariant formalism to investigate the mixed state entanglement structure of time-dependent boosted subsystems in $\TTbar$ deformed CFT$_2$s through the reflected entropy. To this end we utilize the conformal perturbation theory to obtain the R\'enyi reflected entropy through the partition function on replica manifold. The correction to the reflected entropy to the first order in the deformation parameter $\mu$ in then obtained in the replica limit for finite temperature and finite sized systems. We verify our field theoretic computations by obtaining the dual EWCS in the corresponding bulk cut-off AdS$_3$ geometries and find perfect agreement between the two.

	\end{abstract}
	
	\clearpage
	
	\tableofcontents
	
	\clearpage	

%

\section{Introduction} \label{sec:Intro}

The AdS/CFT correspondence \cite{Maldacena:1997re,Gubser:1998bc} provides a non-perturbative framework for gravitational physics in asymptotically anti de Sitter (AdS) spacetime in terms of conformal field theories (CFT) at the asymptotic boundary of the spacetime. Recent advancements have revealed that quantum entanglement plays a central role in the precise mechanism for the emergence of bulk spacetime \cite{VanRaamsdonk:2010pw,Maldacena:2013xja}. The inception of this concept traces back to the holographic characterization of entanglement entropy using the Ryu-Takayanagi (RT) prescription \cite{Ryu:2006bv,Ryu:2006ef}, which subsequently paved the way for significant progress in this field \cite{0705.0016,1102.0440,1304.4926,1607.07506}. 

However, it is well known in quantum information theory that for bipartite mixed states, entanglement entropy receives contributions from irrelevant classical and quantum correlations, and thus fails to be viable measure of quantum entanglement. To address the characterization of mixed state entanglement, several computable entanglement and correlation measures have been introduced in the literature, including entanglement negativity \cite{Vidal:2002zz}, entanglement of purification \cite{Takayanagi:2017knl}, reflected entropy \cite{Dutta:2019gen}, balanced partial entanglement \cite{Wen:2021qgx}, and odd entanglement entropy \cite{Tamaoka:2018ned}, to name a few. In this work, we focus on the reflected entropy, a bipartite mixed state correlation measure associated with the canonical purification of a given mixed state in a doubled Hilbert space \cite{Engelhardt:2017aux,Engelhardt:2018kcs,Dutta:2019gen}. Interestingly, the reflected entropy was posited as the holographic dual to the minimal cross section of the \textit{entanglement wedge}, the bulk codimension-one region bounded by a subsystem and its RT surface \cite{Czech:2012bh,Wall:2012uf,Headrick:2014cta}.

In a separate context, introducing deformations in the asymptotic region of the spacetime corresponds to the inclusion of irrelevant deformations in the dual CFT. However, managing such deformations poses considerable challenges, with only limited exceptions where they can be effectively addressed. In this connection, a class of CFT$_2$s where the conformal symmetry is disturbed through an irrelevant deformation introduced through the determinant of the stress tensor, has been a topic of interest in recent past. This class of CFTs have generally been termed as $\TTbar$ deformed CFTs forming a one-parameter generalization of the original (undeformed) CFT$_2$. It has been shown that these theories retain their integrability and are solvable as the energy spectrum, partition function and S-matrix are exactly determinable \cite{Zamolodchikov:2004ce, Cavaglia:2016oda, Smirnov:2016lqw}. Furthermore, a holographic dual of these $\TTbar$ deformed CFTs was forwarded in \cite{McGough:2016lol} in terms of AdS geometries with the asymptotic boundary situated at finite radial distance. More precisely, the AdS geometry is identical to the one that is dual to the undeformed CFT, except that the cut-off surface is now pushed further into the bulk. The authors in \cite{McGough:2016lol} also demonstrated the matching of the two-point function, the energy spectrum and the partition function of the deformed CFT with the holographic computations, providing substantiation for their proposal.\footnote{See \cite{1711.02690, 1707.08118, 1801.02714, 1801.09708, 1805.10287, 1807.11401, 1808.07760, 1902.10893, 1707.05800} for further developments in this direction.} The entanglement structure of pure and mixed states have further been investigated through conformal perturbation theory in these field theories \cite{Chen:2018eqk, 1806.07444, 1909.13808, 1904.00716, 1906.03894, 1904.04408, 1812.00545, 1911.04618, 1907.12603, 1908.10372,Asrat:2020uib,Basu:2023aqz,Basu:2023bov,2302.13872}.

In this article, we address the the issue of mixed state entanglement in $\TTbar$ deformed CFT$_2$s through the reflected entropy\footnote{Note that, in \cite{Asrat:2020uib}, the authors have also computed the reflected entropy in theories with $\TTbar$  and $\textrm{J}\bar{\textrm{T}}$ deformations for time-independent setups featuring symmetrically placed subsystems.}. In particular we construct a covariant formalism to obtain the reflected entropy for boosted subsystems in such theories. To this end, we perform a conformal perturbative analysis by requiring that the deformation parameter $\mu \ll 1$. In this limit, we obtain the R\'enyi reflected entropy through the partition function on the $nm$-sheeted replica manifold which admits a perturbative expansion in $\mu$, with the leading order contribution arising from the original (unperturbed) CFT$_2$. The correction to the reflected entropy is then obtained to the first order in $\mu$ for such time-dependent boosted subsystems in finite temperature and finite-sized systems. Furthermore, by utilizing the relation between the deformation parameter $\mu$ with the location of this new cut-off surface $r_c$ of the dual AdS$_3$ geometry, we obtain the bulk EWCS in the dual cut-off geometries described by BTZ black holes and global AdS$_3$ spacetimes. We find exact agreement for the first order correction between the bulk and the field theory computations in the large central charge limit, extending the holographic duality of the reflected entropy with the bulk EWCS. Furthermore, we investigate the nature of timelike entanglement, a concept recently introduced in \cite{2210.09457,2210.12963,2302.11695}, in such CFT$_2$s with $\TTbar$ deformation. In particular, we obtain the leading order corrections to the reflected entropy between timelike subsystems in a $\TTbar$ deformed CFT$_2$ via analytic continuation of our results for boosted spacelike subsystems. Although a suitable geometric construction for the bulk entanglement wedge for multiple subregions remains elusive, a naive analytic continuation of the spacelike EWCS matches identically with our field theoretic computations.

The rest of the article is organized as follows. In \cref{sec:review}, we review the salient features of the $\TTbar$ deformed CFT$_2$s and the dual cut-off AdS$_3$ geometries. We also briefly review the salient features of the reflected entropy and its holographic dual, the entanglement wedge cross section. In \cref{sec:SR}, we compute the reflected entropy for disjoint, adjacent and single boosted subsystems in finite temperature and finite-sized $\TTbar$ deformed CFT$_2$s. We also perform an analytic continuation of these results to timelike subsystems. In \cref{sec:EW}, we perform the corresponding holographic computation for the bulk EWCS in cut-off AdS$_3$ geometries for spacelike and timelike subsystems. Finally in \cref{sec:summary}, we summarize our work and draw conclusions.

\section{Review of earlier literature} \label{sec:review}
\subsection{$\TTbar$ deformed CFTs} \label{sec:TTbar-review}

We begin by briefly reviewing the class of theories deformed by the double-trace composite operator \cite{Zamolodchikov:2004ce}
\begin{equation} \label{TTbar-operator}
	\left(\TTbar\right)\equiv \frac{1}{8} \left( T_{ab}\,T^{ab} -\left(T_{a}^a\right) ^2 \right) \,,
\end{equation}
which are termed as $\TTbar$ deformed theories in the literature. These theories form a one-parameter generalization of CFT$_2$s characterized by the deformation parameter $\mu$. The action of the $\TTbar$ deformed theories $S_\text{QFT}^{(\mu)}$ may be characterized by the following flow equation,
\begin{equation} \label{flow-equation}
	\frac{\text{d} S_\text{QFT}^{(\mu)}}{\text{d} \mu} = \int \text{d}^2 w \, ( \TTbar )_\mu \, \qquad , \qquad \qquad S_\text{QFT}^{(\mu)}\Big\vert_{\mu = 0} = S_\text{CFT}
\end{equation}
where $\TTbar$ is constructed out of the stress tensor of the deformed theory, $S_\text{CFT}$ is the action of the undeformed CFT, and $\mu \geq 0$ is considered to be non-negative. If the deformation parameter is considered to be small $\mu \ll 1$, the action $S_\text{QFT}^{(\mu)}$ may be expanded perturbatively to the first order as follows
\begin{equation} \label{TTbar-def}
	S_\text{QFT}^{(\mu)} = S_\text{CFT} + \mu \int_{\mathcal{M}} \text{d}^2 w \, T \bar T \,,
\end{equation}
where $T \equiv T_{ww}$, $\bar{T} \equiv \bar{T}_{\bar{w} \bar{w}} $ are the components of the stress tensor of the \textit{undeformed} theory in the complex $(w, \bar{w})$ plane. Note that, $T^a_a$ has been omitted in the above expression, as on a flat manifold, any correlation function involving an insertion of $T^a_a$ always results in a vanishing contribution. All our computations in this thesis will be performed on a cylindrical manifold, justifying the above omission. Additionally, we will be working within the perturbative regime of the deformation parameter throughout this thesis.


\subsection{Cut-off AdS geometries} \label{sec:cut-off-AdS-review}
\begin{figure}[h!]
	\centering
	\includegraphics[scale=0.65]{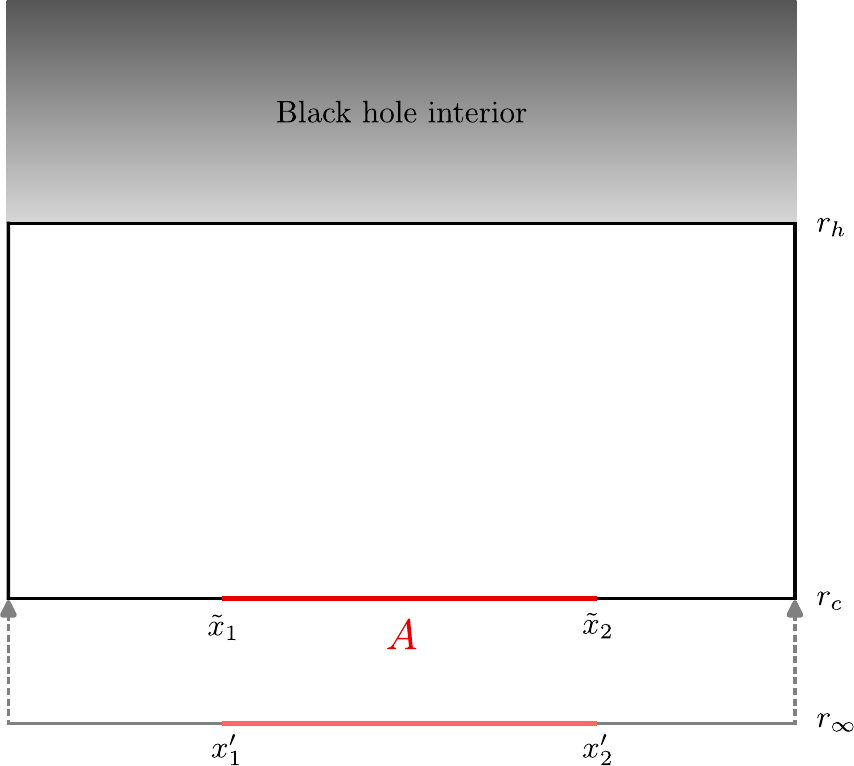}
	\caption{Schematics of the cut-off BTZ geometry dual of $\TTbar$ deformed CFT$_2$ at a finite temperature. The new cut-off surface is placed at $r_c$. The coordinates of a subsystem $A \equiv [x'_1, x'_2]$ in CFT are scaled to $[\tilde{x}_1, \tilde{x}_2]$ while pushing the cut-off surface inside the bulk. Figure modified from \cite{Asrat:2020uib}.}
	\label{fig:ttbar-bulk}
\end{figure}

For these deformed CFT$_2$s with $\mu > 0$, the holographic dual has been proposed to be described by asymptotically AdS$_3$ geometries with asymptotic boundary pushed deeper into the bulk \cite{McGough:2016lol}. In particular, the holographic dual to $T \bar T$ deformed CFT$_2$s are characterized by the AdS geometry identical to the one that is dual to the undeformed CFT, except for the cut-off surface, which satisfies Dirichlet boundary conditions at a finite radial distance:
\begin{align}
	\text{d}s^2=R^2\left(\frac{\text{d}r^2}{r^2}+r^2 g_{ab}\text{d}x^a\text{d}x^b\right)~~,~~
	r<r_c=\sqrt{\frac{6R^4}{\pi c \mu}}=\frac{R^2}{\epsilon_c}\,,\label{rc-generic}
\end{align}
where $R$ is the AdS$_3$ radius, $c$ is the central charge of the deformed CFT$_2$ and the subscript $c$ on the UV cut-off $\epsilon_c$ signifies that the CFT$_2$ defined on the cut-off surface is pushed inside the bulk. An example of this proposal for $\TTbar$ deformed CFT$_2$ at a finite temperature dual to cut-off BTZ black hole, is depicted in \cref{fig:ttbar-bulk}. The above holographic proposal has passed several tests including the explanation of the light-cone deformations, reproducing the energy spectrum, a geometric description of the exact RG flow \cite{McGough:2016lol} as well as holographic characterizations of entanglement and correlation \cite{Chen:2018eqk, Jeong:2019ylz, Asrat:2020uib,Basu:2023bov,Basu:2023aqz}.

\subsection{Reflected entropy} \label{sec:SR-review}

In this subsection, we briefly review the mixed state correlation measure introduced in \cite{Dutta:2019gen} and termed as the reflected entropy. To this end, we begin with the bipartite mixed state $A \cup B$ described by the density matrix $\rho_{AB}$. The canonical purification of $\rho_{AB}$ is given by the state $\ket{\sqrt{\rho_{AB}}}$ defined on the doubled Hilbert space $\mathcal{H}_{A} \otimes \mathcal{H}_{B} \otimes \mathcal{H}_{A^\star} \otimes \mathcal{H}_{B^\star}$ where $A^\star$ and $B^\star$ are the CPT conjugate copies of the subsystems $A$ and $B$ respectively. The reflected entropy $S_R$ between the subsystems $A$ and $B$ is then defined as the von Neumann entropy of the reduced density matrix $\rho_{A A^{\star}} = \Tr_{B B^\star} \ket{\sqrt{\rho_{AB}}}\bra{\sqrt{\rho_{AB}}}$ as follows, 
\begin{align}\label{SR-QIT}
	S_R(A: B) = S_{vN} (\rho_{AA^\star})_{\sqrt{\rho_{AB}}}\,.
\end{align}

A replica technique to obtain the reflected entropy for bipartite mixed states in CFT$_2$s was also developed in \cite{Dutta:2019gen}. This involved the construction of the state $\ket{\psi_m} \equiv \ket{\rho_{AB}^{m/2}}$ on an $m$-sheeted replica manifold with $m \in 2 \mathbb{Z}^+$. This state $\ket{\psi_m}$ may be understood as the purification of $\rho_{AB}^m$. R\'enyi reflected entropy $S_n (A A^\star)_{\psi_m}$ may then be defined as the R\'enyi (entanglement) entropy
corresponding to the reduced density matrix $\rho^{(m)}_{A A^\star}= \Tr_{B B^\star} \ket{\psi_{m}} \bra{\psi_{m}}$. Note here that, unlike the R\'enyi entropy, the R\'enyi reflected entropy is a two-parameter generalization of the reflected entropy where the parameter $m$ is required to construct the pure state $\ket{\psi_m}$ and $n$ is the usual R\'enyi index. The reflected entropy is obtained through the analytic continuation of both these parameters to unity as\footnote{The two replica limits $n \to 1$ and $m \to 1$ do not commute for all cases, as discussed in \cite{Akers:2021pvd, Akers:2022max, Kusuki:2019evw}. However, in this article, the two limits may be implemented interchangeably without any difference in the results.}
\begin{equation} \label{SR-def}
	S_{R} (A : B) = \lim_{n, m \to1} S_{n} (A A^\star)_{\psi_{m}}.
\end{equation}
For any generic configuration, the R\'enyi reflected entropy may be obtained through the partition functions $\zbar_{n,m}$ on the $nm$-replicated and $\zbar_{1,m}$ on the $m$-replicated manifold as follows
\begin{equation} \label{renyi-def}
	\begin{aligned}
		S_n \left(A A^\star \right)_{\psi_m} = \frac{1}{1-n} \log \frac{\zbar_{n,m}}{\zbar_{1,m}^n} \,.
	\end{aligned}
\end{equation}


The authors in \cite{Dutta:2019gen} also established a holographic duality for the reflected entropy in terms of the bulk entanglement wedge cross section. In particular they showed that the reflected entropy is described holographically by twice the bulk entanglement wedge cross section (EWCS). In the next subsection, we provide a brief review of this bulk EWCS.

\subsection{Entanglement wedge cross section} \label{sec:EW-review}
The entanglement wedge is a codimension-one bulk region dual to a density matrix. For a bipartite mixed state $\rho_{AB}$, the entanglement wedge $\chi_{AB}$ is defined as the bulk region bounded by the subsystem $A\cup B$ and the extremal (minimal) surface $\Gamma_{AB}$ homologous to the subsystem computing the entanglement entropy. The minimal cross-section of the entanglement wedge (EWCS) serves as a natural candidate to quantify mixed state entanglement and correlations corresponding to a reduced density matrix\footnote{Note that, several other entanglement and correlation measures such as the entanglement negativity, entanglement of purification, odd entanglement entropy and the balanced partial entanglement have been proposed as putative duals of the entanglement wedge cross section (EWCS). However, the most promising of these dualities stand out to be of that with the reflected entropy established in \cite{Dutta:2019gen} through a gravitational path integral formalism utilizing the explicit holographic construction discussed in \cite{Engelhardt:2017aux}.}. 

In the AdS$_3$/CFT$_2$ setup the extremal curves are always given by geodesic segments and the EWCS may be computed comprehensively in the embedding coordinate formalism where one embeds the AdS$_3$ geometry in $\mathbb{R}^{2,2}$ as follows
\begin{align}
	\text{d}s^2=\eta_{\mu\nu}\text{d}X^{\mu}\text{d}X^{\nu}~~,~~X^{\mu}X_{\mu}=-R^2\,.\label{R2,2-metric}
\end{align}
Then, the EWCS between two disjoint subsystems $A=[X_1,X_2]$ and $B=[X_3,X_4]$ may be computed through \cite{Kusuki:2019evw}
\begin{align}
	E_W(A:B)=\frac{1}{4G_N}\cosh^{-1}\left(\frac{1+\sqrt{u}}{\sqrt{v}}\right)\,,\label{EW-embedding-disj}
\end{align}
where $u$ and $v$ are defined, with $\zeta_{ij}=-X_i.X_j$, as follows
\begin{align}
	u=\frac{\zeta_{12}\zeta_{34}}{\zeta_{13}\zeta_{24}}~~,~~v=\frac{\zeta_{14}\zeta_{23}}{\zeta_{13}\zeta_{24}}\,.
\end{align}
In a similar manner, for two adjacent subsystems $A=[X_1,X_2]$ and $B=[X_2,X_3]$ the EWCS is given by \cite{Basu:2023jtf}
\begin{align}
	E_W(A:B)=\frac{1}{4G_N}\cosh^{-1}\left(\sqrt{\frac{2\zeta_{12}\zeta_{23}}{\zeta_{13}}}\right)\,.\label{EW-embedding-adj}
\end{align}

\section{Reflected entropy in $\TTbar$ deformed CFTs} \label{sec:SR}
In this section, we construct a suitable replica technique to obtain the reflected entropy in CFT$_2$s perturbed with a $\TTbar$ operator. To this end, we begin by considering any generic bipartite mixed state in such a $\TTbar$ deformed CFT$_2$ defined on a manifold $\mathcal{M}$. As discussed in \cref{sec:SR-review}, to obtain the reflected entropy in a CFT$_2$ it is required to consider an $nm$-sheeted Riemannian manifold $\mathcal{M}_{nm}$, constructed through the replication of the original manifold $\mathcal{M}$. The partition function on such a replicated manifold may be written as
\begin{equation}
	{\zbar}_{n,m} = \int_{\mathcal{M}_{nm}} \mathcal{D}\Phi \,e^{-S^{(\mu)}_\text{QFT}[\Phi]},
\end{equation}
where $S^{(\mu)}_\text{QFT}$ is the action for the $\TTbar$ deformed CFT$_2$ on the replicated manifold $\mathcal{M}_{nm}$. Considering the $\TTbar$ deformation to be perturbative by taking $\mu \ll 1$ in \cref{TTbar-def}, we may obtain the following ratio of the partition functions on the $nm$-sheeted replica manifold and the $m$-sheeted replica manifold,
\begin{equation}\label{partition-ratio}
	\frac{{\zbar}_{n,m}}{{\zbar}_{1,m}^n} = \frac{\int_{\mathcal{M}_{nm}} e^{-S_\text{CFT}}}{\left[ \int_{\mathcal{M}_{m}} e^{-S_\text{CFT}}\right]^n} \left( 1 - \mu \int_{\mathcal{M}_{nm}} \langle \ttbar \rangle_{\mathcal{M}_{nm}} + n \mu \int_{\mathcal{M}_{m}} \langle \ttbar \rangle_{\mathcal{M}_{m}} + \mathcal{O} (\mu^2) \right) \, ,
\end{equation}
where the first term gives the leading order result arising from the original (unperturbed) field theory. Here the expectation value of the $\ttbar$ operator on $\mathcal{M}_{nm}$ may be determined by employing the twist operators for any general bipartite configuration as follows
\begin{equation} \label{TTbar-exp}
	\begin{aligned}
		\int_{\mathcal{M}_{nm}} \langle \ttbar \rangle_{\mathcal{M}_{nm}} =& \sum_{k=1}^{nm} \int_\mathcal{M} \frac{\langle T_k (w) \bar T_k (\bar w) \Pi_i \sigma_i  (w_i,\bar w_i) \rangle_\mathcal{M}}{\langle \Pi_i \sigma_i  (w_i,\bar w_i) \rangle_\mathcal{M}}\\
		=& \int_\mathcal{M} \frac{1}{nm} \frac{\langle T^{(nm)} (w) \bar T^{(nm)} (\bar w) \Pi_i \sigma_i  (w_i,\bar w_i) \rangle_\mathcal{M}}{\langle \Pi_i \sigma_i  (w_i,\bar w_i) \rangle_\mathcal{M}} \,,
	\end{aligned}
\end{equation}
where $T_k$ are defined on the $k$-th replica sheet and $T^{(nm)}$ correspond to the total energy momentum tensor on the $nm$-replicated manifold. The expectation value of the $\ttbar$ operator on the replicated manifold $\mathcal{M}_{m}$ may also be determined in a similar fashion. 

The correction to the R\'enyi reflected entropy may then be identified by utilizing \cref{partition-ratio} in \cref{renyi-def} to be
\begin{equation} \label{Snm-ttbar-def}
	\delta S_{n,m} (A A^*) = \frac{\mu}{n-1} \left(\int_{\mathcal{M}_{nm}}\langle \ttbar \rangle_{\mathcal{M}_{nm}} - n \int_{\mathcal{M}_{m}} \langle \ttbar \rangle_{\mathcal{M}_{m}} \right) \,,
\end{equation}
which in replica limit $n,m \to 1$, will lead to the correction in the reflected entropy due to such $\TTbar$ deformation in CFT$_2$s.

\subsection{Thermal CFT$_2$s with $\TTbar$ deformation} \label{sec:SR-beta}
In this subsection, we consider the $\TTbar$ deformation of a CFT$_2$ at a finite temperature. Thus the manifold $\mathcal{M}$ in this case is described by a thermal cylinder with a circumference given by the inverse temperature $\beta$. We may transform this cylinder $\mathcal{M}$ described by coordinates $(w, \bar w)$ to a complex plane $\mathbb{C}$ with coordinates $(z, \bar z)$ by utilizing the following exponential map,
\begin{equation} \label{cyl-map-beta}
	z = e^{\frac{2 \pi w}{\beta}} \qquad , \qquad \bar z = e^{\frac{2 \pi \bar w}{\beta}} \,.
\end{equation}
It is now possible to compute the expectation value in \cref{TTbar-exp} by transforming the correlation functions to the complex plane and by utilizing the conformal Ward identities in the numerator \cite{Jeong:2019ylz,Asrat:2020uib,Basu:2023bov,Basu:2023aqz}. 

In the following, we consider various bipartite mixed states in such finite temperature CFT$_2$s and obtain the correction to the reflected entropy. Subsequently we consider an analytic continuation of our results to obtain the correction for purely timelike subsystems.

\subsubsection{Two disjoint subsystems} \label{sec:SR-beta-disj}
We begin by considering two disjoint subsystems described by $A \equiv [(w_1,\bar w_1),(w_2,\bar w_2)]$ and $B \equiv [(w_3,\bar w_3),(w_4,\bar w_4)]$ with $w_k = x_k + i \tau_k$ in a thermal CFT$_2$ with a $\TTbar$ deformation. Computing the expectation value in \cref{TTbar-exp} via the map \eqref{cyl-map-beta} and substituting it in \cref{Snm-ttbar-def} lead to the following correction in the R\'enyi reflected entropy for this configuration,
\begin{equation} \label{Snm-disj-beta}
	\begin{aligned}
		\delta S_n \left(A A^\star \right)_{\psi_m} = \frac{\mu}{n-1}\int_\mathcal{M} &\Bigg[-\frac{2 \pi^4 c}{3 \beta^4} \Bigg(z^2 \sum_{i=1}^{4} \left( \frac{{h_{g^{}_i}}-n\, {h_{g^{}_m}}}{(z-{z_i})^2} + \frac{\partial_{z_i}({\log \langle \sigma \rangle}-{n \, \log \langle \sigma_m \rangle})}{z-{z_i}} \right)\\
		&\qquad \qquad \quad +{\bar{z}}^2 \sum_{i=1}^{4} \left( \frac{{\bar{h}_{g^{}_i}}-n\, {\bar{h}_{g^{}_m}}}{(\bar{z}-{\bar{z}_i})^2} + \frac{\partial_{\bar{z}_i}({\log \langle \sigma \rangle}-{n \, \log \langle \sigma_m \rangle})}{\bar{z}-{\bar{z}_i}} \right) \Bigg)\\
		&+\frac{16 \pi^2}{\beta^2 \, m} \Bigg(\frac{z^2 {\bar{z}}^2}{n} \sum_{i,j=1}^{4} \left( \frac{{h_{g^{}_i}}}{(z-{z_i})^2} + \frac{\partial_{z_i}{\log \langle \sigma \rangle}}{z-{z_i}} \right) \left( \frac{{\bar{h}_{g^{}_j}}}{(\bar{z}-{\bar{z}_j})^2} + \frac{\partial_{\bar{z}_j}{\log \langle \sigma \rangle}}{\bar{z}-{\bar{z}_j}} \right)\\
		& -n \, z^2 {\bar{z}}^2 \sum_{i,j=1}^{4} \left( \frac{{h_{g^{}_m}}}{(z-{z_i})^2} + \frac{\partial_{z_i}{\log \langle \sigma_m \rangle}}{z-{z_i}} \right) \left( \frac{{\bar{h}_{g^{}_m}}}{(\bar{z}-{\bar{z}_j})^2} + \frac{\partial_{\bar{z}_j}{\log \langle \sigma_m \rangle}}{\bar{z}-{\bar{z}_j}}\right)\Bigg)\Bigg].
	\end{aligned}
\end{equation}
In the above, we have defined
\begin{align}
	\log \langle \sigma \rangle &\equiv \log \langle \sigma_{g^{}_A} (z_1,\bar{z}_1) \sigma_{g^{-1}_A} (z_2,\bar{z}_2) \sigma_{g^{}_B} (z_3,\bar{z}_3) \sigma_{g^{-1}_B} (z_4,\bar{z}_4) \rangle  \notag \\
	&\approx h_{g^{-1}_A g^{}_B} \log \frac{1 + \sqrt{\eta}}{1 - \sqrt{\eta}} + \bar{h}_{g^{-1}_A g^{}_B} \log \frac{1 + \sqrt{\bar{\eta}}}{1 - \sqrt{\bar{\eta}}} \,, \label{log-sigma-disj-beta}\\
	\log \langle \sigma_m \rangle &\equiv \log \langle \sigma_{g^{}_m} (z_1,\bar{z}_1) \sigma_{g^{-1}_m} (z_2,\bar{z}_2) \sigma_{g^{}_m} (z_3,\bar{z}_3) \sigma_{g^{-1}_m} (z_4,\bar{z}_4) \rangle = \lim_{n \to 1} \log \langle \sigma \rangle \,, \label{log-sigma-m-disj-beta}
\end{align}
where the cross-ratio $\eta=\frac{z_{12} z_{34}}{z_{13} z_{24}}$ with $z_{ij} = z_i - z_j$. Also $h_{g_i} = \bar{h}_{g_i}$, $h_{g_m} = \bar{h}_{g_m}$ and $h_{g^{-1}_A g^{}_B} = \bar{h}_{g^{-1}_A g^{}_B}$, respectively represent the conformal dimensions of the twist operators in \cref{log-sigma-disj-beta} at position $z_i$, of $\sigma_{g_m}$ in \cref{log-sigma-m-disj-beta} and of the composite twist operator $\sigma_{g^{-1}_A g^{}_B}$, and have the following form
\begin{equation}\label{conformal-dimensions}
	\begin{aligned}
		h_{g_i} = n \, h_{g_m} = \frac{n \, c}{24}\left( m - \frac{1}{m} \right) ~~~,~~~ h_{g^{-1}_A g^{}_B} = \frac{c}{12}\left( n - \frac{1}{n} \right) \, .
	\end{aligned}
\end{equation}
Note that the conformal dimension of $\sigma_{g^{-1}_A g^{}_B}$ appear in \cref{log-sigma-disj-beta} as this operator provides the dominant contribution in the conformal block expansion of the corresponding four-point twist correlator \cite{Dutta:2019gen}.

On simplifying \cref{Snm-disj-beta}, we may obtain the following integral,
\begin{equation} \label{Snm-disj-beta-int}
	\delta S_n \left(A A^\star \right)_{\psi_m} = - \int_\mathcal{M} \text{d}^2 w \frac{\pi^4 c^2 \mu}{9 \beta^4} \left( \frac{z^2 \sqrt{(z_1-z_2) (z_1-z_3) (z_2-z_4) (z_3-z_4)}}{(z-z_1) (z-z_2) (z-z_3) (z-z_4)}+ \text{c.c.} \right)  \,,
\end{equation}
where $z = e^{\frac{2 \pi (x + i \tau)}{\beta}}$ and $\text{c.c.}$ represent the complex conjugate of the first term. This integral may easily be solved following the techniques described in \cite{Jeong:2019ylz} where similar integrals appear in the computation of correction to the entanglement entropy. Finally, in the replica limit $n,m \to 1$, the correction to the reflected entropy for the given configuration of two disjoint subsystems may be obtained as follows
\begin{equation} \label{SR-disj-beta}
	\begin{aligned}
		\delta S_R = &\frac{\pi ^4 c^2 \mu }{18 \beta ^3 \sqrt{\eta }} \left(\mathcal{P}^\beta_{134}-\mathcal{P}^\beta_{234}+\mathcal{P}^\beta_{312}-\mathcal{P}^\beta_{412}\right)\\
		&-\frac{\pi ^4 c^2 \mu}{18 \beta ^3 \sqrt{{\bar\eta}}} \left(\bar{\mathcal{P}}^\beta_{134}-\bar{\mathcal{P}}^\beta_{234}+\bar{\mathcal{P}}^\beta_{312}-\bar{\mathcal{P}}^\beta_{412}\right)\,,
	\end{aligned}
\end{equation}
where we have defined 
\begin{equation} \label{Pijk-beta}
	\begin{aligned}
		\mathcal{P}^\beta_{ijk}&=\frac{x_i \sinh \left(\frac{\pi ( t_{jk} + x_{jk} )}{\beta }\right)}{\sinh \left(\frac{\pi ( t_{ij} +x_{ij} )}{\beta }\right) \sinh \left(\frac{\pi ( t_{ik} +x_{ik} )}{\beta }\right)} \,, \\
		\bar{\mathcal{P}}^\beta_{ijk}&=\frac{x_i \sinh \left(\frac{\pi ( t_{jk} - x_{jk} )}{\beta }\right)}{\sinh \left(\frac{\pi ( t_{ij} -x_{ij} )}{\beta }\right) \sinh \left(\frac{\pi ( t_{ik} -x_{ik} )}{\beta }\right)}\,.
	\end{aligned}
\end{equation}
We have also analytically continued to the real time $t$ to explicitly observe the time dependence of our result, and the finite temperature cross-ratios $\eta$ and $\bar \eta$ are given by
\begin{equation}\label{cross-ratio-generic}
	\begin{aligned}
		\eta = \frac{\sinh \left(\frac{\pi(t_{12} + x_{12})}{\beta}\right) \sinh \left(\frac{\pi(t_{34} + x_{34})}{\beta}\right)}{\sinh \left(\frac{\pi ( t_{13} + x_{13} )}{\beta}\right) \sinh \left(\frac{\pi ( t_{24} + x_{24} )}{\beta}\right)} \,, \\
		\bar \eta =\frac{\sinh \left(\frac{\pi(t_{12} - x_{12})}{\beta}\right) \sinh \left(\frac{\pi(t_{34} - x_{34})}{\beta}\right)}{\sinh \left(\frac{\pi ( t_{13} - x_{13} )}{\beta}\right) \sinh \left(\frac{\pi ( t_{24} - x_{24} )}{\beta}\right)}\,.
	\end{aligned}
\end{equation}
It may be checked here that the correction to the leading order in \cref{SR-disj-beta} is negative for spacelike subsystems, indicative of a reduction in the entanglement between the two parties forming the bipartite mixed state. 

\subsubsection*{Timelike entanglement} \label{sec:SR-beta-disj-T}
We now investigate the timelike entanglement structure of the bipartite mixed state of two disjoint subsystems through an analytic continuation of the above correction to the reflected entropy. We consider purely timelike subsystems $A \equiv [(x,t_1), (x,t_2)]$ and $B \equiv [(x,t_3), (x,t_4)]$ on a thermal cylinder describing the CFT$_2$ with a $\TTbar$ deformation and substitute these in \cref{SR-disj-beta} to obtaining a vanishing correction. Thus the timelike reflected entropy $S_{R}^{\text{(T)}}$ for this configuration is same as the one with no $\TTbar$ deformation and may be obtained through the analytic continuation of the result obtained in \cite{Afrasiar:2022wzn} as follows
\begin{equation} \label{SR-disj-beta-T}
	\begin{aligned}
		S_{R}^\text{(T)} = &\frac{c}{3} \log \frac{1 + \sqrt{\frac{\sinh \left(\frac{\pi t_{12}}{\beta}\right) \sinh \left(\frac{\pi t_{34}}{\beta}\right)}{\sinh \left(\frac{\pi t_{13}}{\beta}\right) \sinh \left(\frac{\pi t_{24}}{\beta}\right)}}}{1 - \sqrt{\frac{\sinh \left(\frac{\pi t_{12}}{\beta}\right) \sinh \left(\frac{\pi t_{34}}{\beta}\right)}{\sinh \left(\frac{\pi t_{13}}{\beta}\right) \sinh \left(\frac{\pi t_{24}}{\beta}\right)}}} \,.
	\end{aligned}
\end{equation}
Notice that, unlike the timelike EE, the timelike reflected entropy for this configuration does not have any imaginary part. 

The above vanishing result is similar to the findings of \cite{Jiang:2023ffu, Basu:2023aqz} where the authors observed no correction to the entanglement entropy and the odd entanglement entropy for a timelike subsystem in the $\TTbar$ deformed thermal CFT$_2$. This was explained, holographically, owing to the orientation of the subsystems with respect to the cylinder.

\subsubsection{Two adjacent subsystems} \label{sec:SR-beta-adj}
We now proceed to case of two adjacent subsystems described by $A \equiv [(w_1,\bar w_1),(w_2,\bar w_2)]$ and $B \equiv [(w_2,\bar w_2),(w_3,\bar w_3)]$ in a thermal CFT$_2$ with $\TTbar$ deformation. The correction to the R\'enyi entropy \eqref{Snm-ttbar-def} for this case may be obtained by computing the expectation value in \cref{TTbar-exp}, as follows
\begin{equation}\label{Snm-adj-beta}
	\begin{aligned}
		\delta S_n \left(A A^\star \right)_{\psi_m} = \frac{\mu }{n-1} &\int_\mathcal{M}\Bigg[-\frac{2 \pi^4 c}{3 \beta^4} \Bigg(\sum_{i=1}^{3} \left( z^2  \left( \frac{{h_{g_i}}}{(z-{z_i})^2} + \frac{\partial_{z_i}{\log \langle \sigma \rangle}}{z-{z_i}} \right) + {\bar{z}}^2  \left( \frac{{\bar{h}_{g_i}}}{(\bar{z}-{\bar{z}_i})^2} + \frac{\partial_{\bar{z}_i}{\log \langle \sigma \rangle}}{\bar{z}-{\bar{z}_i}} \right) \right)\\
		& -\sum_{i=1,3} \left( n \, z^2 \left( \frac{{h_{g_m}}}{(z-{z_i})^2} + \frac{\partial_{z_i}{\log \langle \sigma_m \rangle}}{z-{z_i}} \right) -n \, {\bar{z}}^2 \left( \frac{{\bar{h}_{g_m}}}{(\bar{z}-{\bar{z}_i})^2} + \frac{\partial_{\bar{z}_i}{\log \langle \sigma_m \rangle}}{\bar{z}-{\bar{z}_i}}\right)\right)\Bigg)\\
		&+\frac{16 \pi^2}{\beta^2 \, m} \Bigg(\frac{z^2 {\bar{z}}^2}{n} \sum_{i,j=1}^{3} \left( \frac{{h_{g_i}}}{(z-{z_i})^2} + \frac{\partial_{z_i}{\log \langle \sigma \rangle}}{z-{z_i}} \right) \left( \frac{{\bar{h}_{g_j}}}{(\bar{z}-{\bar{z}_j})^2} + \frac{\partial_{\bar{z}_j}{\log \langle \sigma \rangle}}{\bar{z}-{\bar{z}_j}} \right)\\
		& -n \, z^2 {\bar{z}}^2 \sum_{i,j=1,3} \left( \frac{{h_{g_m}}}{(z-{z_i})^2} + \frac{\partial_{z_i}{\log \langle \sigma_m \rangle}}{z-{z_i}} \right) \left( \frac{{\bar{h}_{g_m}}}{(\bar{z}-{\bar{z}_j})^2} + \frac{\partial_{\bar{z}_j}{\log \langle \sigma_m \rangle}}{\bar{z}-{\bar{z}_j}}\right)\Bigg)\Bigg]\,,
	\end{aligned}
\end{equation}
where 
\begin{align}
	\log \langle \sigma \rangle & \equiv \log \langle \sigma_{g^{}_A} (z_1,\bar{z}_1) \sigma_{g^{-1}_A g^{}_B} (z_2,\bar{z}_2) \sigma_{g^{-1}_B} (z_3,\bar{z}_3) \rangle \label{log-sigma-adj}\,,\\
	\log \langle \sigma_m \rangle & \equiv \log \langle \sigma_{g^{}_m} (z_1,\bar{z}_1) \sigma_{g_m^{-1}} (z_3,\bar{z}_3) \rangle\,,\label{log-sigma-m-adj}
\end{align}
with conformal dimensions of the twist operators given in \cref{conformal-dimensions}. Simplifying \cref{Snm-adj-beta}, we may arrive at the following integral on the thermal cylinder $\mathcal{M}$,
\begin{equation} \label{Snm-adj-beta-int}
	\begin{aligned}
		\delta S_n \left(A A^\star \right)_{\psi_m} = - \int_\mathcal{M} \text{d}^2 w \frac{\pi^4 c^2 \mu}{9 \beta^4}\left( \frac{z^2 (z_2 - z_1) (z_2 - z_3)}{(z - z_1) (z - z_2)^2 (z - z_3)} + \text{c.c.} \right)\,.
	\end{aligned}
\end{equation}
This integral may again be computed following the techniques described in \cite{Jeong:2019ylz}, which in replica limit $n,m \to 1$ provide the following correction to the reflected entropy,
\begin{equation}
	\begin{aligned}\label{SR-adj-beta}
		\delta S_R = -\frac{\pi^4 c^2 \mu }{9 \beta ^3} \vast(\frac{x_{12} \sinh \left(\frac{2 \pi x_{12}}{\beta}\right)}{\cosh \left(\frac{2 \pi t_{12}}{\beta }\right)-\cosh \left(\frac{2 \pi x_{12}}{\beta }\right)}&+\frac{x_{23}\sinh \left(\frac{2 \pi x_{23}}{\beta }\right)}{\cosh \left(\frac{2 \pi t_{23}}{\beta }\right)-\cosh \left(\frac{2 \pi x_{23}}{\beta }\right)}\\
		&-\frac{x_{13} \sinh \left(\frac{2 \pi x_{13}}{\beta}\right)}{\cosh \left(\frac{2 \pi t_{13}}{\beta}\right)-\cosh \left(\frac{2 \pi x_{13}}{\beta }\right)}\vast)\,,
	\end{aligned}
\end{equation}
where we have analytically continued to the real time $t$. Again, the above correction is always negative for boosted spacelike subsystems hinting towards a reduction in the entangled degrees of freedom shared between the two adjacent subsystems due to the $\TTbar$ deformation.

\subsubsection*{Timelike entanglement} \label{sec:SR-beta-adj-T}
We now consider the mixed state of two adjacent (purely) timelike subsystems $A \equiv [(x,t_1), (x,t_2)]$ and $B \equiv [(x,t_2), (x,t_3)]$ on the thermal cylinder $\mathcal{M}$ describing the CFT$_2$ with a $\TTbar$ deformation. Considering the analytic continuation of \cref{SR-adj-beta}, we again observe a vanishing correction to the reflected entropy in the leading order for the given timelike subsystems. Thus the reflected entropy for this case is simply obtained by analytically continuing the corresponding result of \cite{Afrasiar:2022wzn} as follows
\begin{equation}\label{SR-adj-beta-timelike}
	\begin{aligned}	
		S_R^\text{(T)} = \frac{c}{3}\log\Bigg[\bigg(\frac{4 \beta}{\pi \epsilon}\bigg)\frac{\sinh{\big(\frac{\pi t_{12}}{\beta}\big)}\sinh{\big(\frac{\pi t_{23}}{\beta}\big)}}{\sinh({\frac{\pi t_{13}}{\beta} })}\Bigg] + i \frac{\pi c}{6}\,.
	\end{aligned}	
\end{equation}
Similar imaginary part has been observed to appear in the entanglement entropy for timelike intervals in CFT$_2$s with and without \cite{Doi:2022iyj, Doi:2023zaf, Jiang:2023ffu} $\TTbar$ deformations as well. In the holographic description they correspond to some timelike geodesics constituting the RT surface for the timelike subsystem \cite{Doi:2022iyj, Doi:2023zaf}.

\subsubsection{A single subsystem} \label{sec:SR-beta-sing}
In this subsection we consider a single spacelike subsystem described by $A \equiv [(0,0) , (\ell,t)]$ in a CFT$_2$ at a finite temperature deformed by the $\TTbar$ operator. Similar to the case for the reflected entropy in undeforrmed CFT$_2$s \cite{Basu:2022nds, Afrasiar:2022wzn}, the appropriate replica technique prescription for this configuration involves the consideration of two auxiliary subsystems $B_1 \equiv [(-L,-T) , (0,0)]$ and $B_2 \equiv [(\ell, t), (L,T)]$ adjacent to the given single subsystem on either sides.\footnote{Similar construction has also been utilized in non-relativistic scenarios for the reflected entropy of a single subsystem at a finite temperature in GCFT$_2$s \cite{Basak:2022cjs}.} This is required due to the presence of infinite branch cuts along the cylinder which involve non-trivial gluing among the replica sheets for the R\'enyi reflected entropy. The reflected entropy is obtained with these subsystems in place, and finally the bipartite limit $B_1 \cup B_2 \equiv B \to A^c$ ($L \to \infty$) is considered to obtain the original configuration of the single subsystem.

For this configuration of the single subsystem sandwiched by the auxiliary subsystems, the correction to the R\'enyi reflected entropy in \cref{Snm-ttbar-def} is explicitly given in terms of the twist operators, upon utilizing the conformal map in  \cref{cyl-map-beta} as follows
\begin{equation}\label{Snm-sing-beta}
	\begin{aligned}
		\delta S_n \left(A A^\star \right)_{\psi_m} = \frac{\mu}{n-1}& \int_\mathcal{M}\Bigg[-\frac{2 \pi^4 c}{3 \beta^4} \Bigg(z^2 \sum_{i=1}^{4} \left( \frac{{h_{g^{}_i}}}{(z-{z_i})^2} + \frac{\partial_{z_i}{\log \langle \sigma \rangle}}{z-{z_i}} \right) + {\bar{z}}^2 \sum_{i=1}^{4} \left( \frac{{\bar{h}_{g^{}_i}}}{(\bar{z}-{\bar{z}_i})^2} + \frac{\partial_{\bar{z}_i}{\log \langle \sigma \rangle}}{\bar{z}-{\bar{z}_i}} \right)\\
		&-n \, z^2 \sum_{i=1,4} \left( \frac{{h_{g^{}_m}}}{(z-{z_i})^2} + \frac{\partial_{z_i}{\log \langle \sigma_m \rangle}}{z-{z_i}} \right) -n \, {\bar{z}}^2 \sum_{i=1,4} \left( \frac{{\bar{h}_{g^{}_m}}}{(\bar{z}-{\bar{z}_i})^2} + \frac{\partial_{\bar{z}_i}{\log \langle \sigma_m \rangle}}{\bar{z}-{\bar{z}_i}}\right) \Bigg)\\
		&+\frac{16 \pi^2}{\beta^2 \, m} \Bigg(\frac{z^2 {\bar{z}}^2}{n} \sum_{i,j=1}^{4} \left( \frac{{h_{g^{}_i}}}{(z-{z_i})^2} + \frac{\partial_{z_i}{\log \langle \sigma \rangle}}{z-{z_i}} \right) \left( \frac{{\bar{h}_{g^{}_j}}}{(\bar{z}-{\bar{z}_j})^2} + \frac{\partial_{\bar{z}_j}{\log \langle \sigma \rangle}}{\bar{z}-{\bar{z}_j}} \right)\\
		&-n \, z^2 {\bar{z}}^2 \sum_{i,j=1,4} \left( \frac{{h_{g^{}_m}}}{(z-{z_i})^2} + \frac{\partial_{z_i}{\log \langle \sigma_m \rangle}}{z-{z_i}} \right) \left( \frac{{\bar{h}_{g^{}_m}}}{(\bar{z}-{\bar{z}_j})^2} + \frac{\partial_{\bar{z}_j}{\log \langle \sigma_m \rangle}}{\bar{z}-{\bar{z}_j}}\right)\Bigg)\Bigg]\,,
	\end{aligned}
\end{equation}
where 
\begin{align} 
	\log \langle \sigma \rangle \equiv& \log \langle \sigma_{g^{}_B} (z_1,\bar{z}_1) \sigma_{g^{-1}_B g^{}_A} (z_2,\bar{z}_2) \sigma_{g^{-1}_A g^{}_B} (z_3,\bar{z}_3) \sigma_{g^{-1}_B} (z_4,\bar{z}_4) \rangle \label{log-sigma-sing}\\
	\log \langle \sigma_m \rangle \equiv &\log \langle \sigma_{g_m} (z_1,\bar{z}_1) \sigma_{g_m^{-1}} (z_4,\bar{z}_4) \rangle = - h_{g_m} \log ( z_4 - z_1 ) - \bar{h}_{g_m} \log ( \bar{z}_4 - \bar{z}_1 )\,. \label{log-sigma-m-sing}
\end{align}
The four-point twist correlator in \cref{log-sigma-sing} is given by \cite{Basu:2022nds, Afrasiar:2022wzn}
\begin{equation}\label{four-point-single}
	\begin{aligned} 
		\langle \sigma_{g^{}_B} (z_1,\bar{z}_1) \sigma_{g^{-1}_B g^{}_A} (z_2,\bar{z}_2) \sigma_{g^{-1}_A g^{}_B} (z_3,\bar{z}_3) \sigma_{g^{-1}_B} (z_4,\bar{z}_4) \rangle = k_{mn} & \left( \frac{1}{z_{14}^{2 h_{g^{}_B}} z_{23}^{2 h_{g^{-1}_A g^{}_B}}} \frac{\mathcal{F}_{mn}(\eta)}{\eta^{h_{g^{-1}_A g^{}_B}}} \right)\\
		\times & \left( \frac{1}{\bar z_{14}^{2 \bar {h}_{g^{}_B}} \bar z_{23}^{2 \bar h_{g^{-1}_A g^{}_B}}} \frac{\mathcal{\bar F}_{mn}(\bar \eta)}{\bar \eta^{\bar h_{g^{-1}_A g^{}_B}}} \right)\,, 
	\end{aligned}
\end{equation}
where $\eta$ and $\bar \eta$ are the cross ratios as defined earlier in $z$ and $\bar z$ coordinates. The non-universal function $\mathcal{F}_{mn} (\eta)$ and $\bar {\mathcal{F}}_{mn} (\bar \eta)$ depend on the full operator content of the theory and in the limit $\eta, \bar \eta \to 1$ and $\eta, \bar \eta \to 0$ may be approximated as \cite{Basu:2022nds}
\begin{equation} \label{non-universal-F}
	\mathcal{F}_{mn} (1) = \bar {\mathcal{F}}_{mn} (1) = 1\,, \qquad \qquad \mathcal{F}_{mn} (0) = \bar {\mathcal{F}}_{mn} (0) = C_{mn}\,,
\end{equation}
where $C_{mn}$ is a non-universal constant. Using the above, the correction to the reflected entropy may be obtained in the replica limit $n, m \to 1$ of the R\'enyi reflected entropy as
\begin{equation}\label{SR-sing-beta}
	\begin{aligned}
		\delta S_R = - \frac{\pi ^4 c^2 \ell \mu }{9 \beta ^3} \Bigg( \coth \left(\frac{\pi (\ell-t)}{\beta }\right)+&\coth \left(\frac{\pi (\ell+t)}{\beta }\right) - 2 \Bigg)\\
		-& e^{\frac{2 \pi (\ell - t)}{\beta}} \frac{f'\left(e^{\frac{2 \pi (\ell - t)}{\beta}}\right)}{f\left(e^{\frac{2 \pi (\ell - t)}{\beta}}\right)} - e^{\frac{2 \pi (\ell + t)}{\beta}} \frac{\bar {f}'\left(e^{\frac{2 \pi (\ell + t)}{\beta}}\right)}{\bar f \left(e^{\frac{2 \pi (\ell + t)}{\beta}}\right)} \,,
	\end{aligned}
\end{equation}
where the bipartite limit $L \to \infty$ has been implemented. We have also defined the non-universal functions
\begin{align} \label{non-universal-f}
	f(\eta) \equiv \lim_{m,n \to 1} \log [\mathcal{F}_{mn} (\eta)] \,, \qquad \qquad \bar f(\bar \eta) \equiv \lim_{m,n \to 1} \log [\bar {\mathcal{F}}_{mn} (\bar \eta)] \,.
\end{align}
Similar to previous cases, the correction to the reflected entropy for this configuration is always negative for spacelike subsystems hinting towards the reduction in the entangled degrees of freedom between the subsystem $A$ and its compliment.

\subsubsection*{Timelike entanglement} \label{sec:SR-beta-sing-T}
We now consider the case where the subsystem $A$ is purely timelike. It is important to note here that the above construction for a spacelike subsystem involving the auxiliary subsystems should not be extended naively to timelike situations as the apparent problem due to infinite branch cuts is absent in such cases. This is due to the fact that the subsystem is now aligned in the compactified direction of the thermal cylinder making the branch cuts along $A^c$ finite. Thus the reflected entropy is obtained through a two-point twist correlator in this case for which the correction in the reflected entropy given by \cref{Snm-ttbar-def} vanishes.

\subsection{Finite-sized CFT$_2$s with $\TTbar$ deformation} \label{sec:SR-L}
In this subsection, we now proceed to the case of $\TTbar$ deformation of finite-sized CFT$_2$s leading to a cylindrical manifold $\mathcal{M}$ with a compactified spatial direction $x \sim x + L$. This spatial cylinder may be mapped to the complex plane $\mathbb{C}$ through the following,
\begin{equation} \label{cyl-map-L}
	z = e^{-\frac{2 \pi i w}{L}} \qquad , \qquad \bar z = e^{\frac{2 \pi i \bar w}{L}} \,.
\end{equation}
Similar to the finite temperature case, we may now compute the correction to the R\'enyi reflected entropy by obtaining the expectation value of the $\TTbar$ operator in \cref{TTbar-exp}.

\subsubsection{Two disjoint subsystems} \label{sec:SR-L-disj}
Consider the bipartite state of two disjoint subsystems described by $A \equiv [(w_1,\bar w_1),(w_2,\bar w_2)]$ and $B \equiv [(w_3,\bar w_3),(w_4,\bar w_4)]$ in a finite-sized CFT$_2$ with a $\TTbar$ deformation. Here $w_k = x_k + i \tau_k$ with $x_k \sim x_k + L$ and $\tau_k$ as the Euclidean time. The correction to the R\'enyi reflected entropy for this case is given by the same integral as in \cref{Snm-disj-beta-int} with $\beta \to i L$. However the planar coordinates $(z, \bar z)$ are now related to the coordinates $(w, \bar w)$ on the spatial cylinder through \cref{cyl-map-L}. Solving the integral and subsequently taking the replica limit $n,m \to 1$ leads to the following correction for the reflected entropy,
\begin{equation} \label{SR-disj-L}
	\begin{aligned}
		\delta S_R = &\frac{\pi ^4 c^2 \mu }{18 L^3 \sqrt{\xi }} \left(\mathcal{P}^L_{134}-\mathcal{P}^L_{234}+\mathcal{P}^L_{312}-\mathcal{P}^L_{412}\right)\\
		&-\frac{\pi ^4 c^2 \mu }{18 L^3 \sqrt{\bar \xi }} \left(\bar{\mathcal{P}}^L_{134}-\bar{\mathcal{P}}^L_{234}+\bar{\mathcal{P}}^L_{312}-\bar{\mathcal{P}}^L_{412}\right)\,,
	\end{aligned}
\end{equation}
where we have again defined the following functions
\begin{equation} \label{Pijk-L}
	\begin{aligned}
		\mathcal{P}^L_{ijk}&=\frac{t_i \sin \left(\frac{\pi ( t_{jk} + x_{jk} )}{L}\right)}{\sin \left(\frac{\pi ( t_{ij} +x_{ij} )}{L}\right) \sin \left(\frac{\pi ( t_{ik} +x_{ik} )}{L}\right)} \,, \\
		\bar{\mathcal{P}}^L_{ijk}&=\frac{t_i \sin \left(\frac{\pi ( t_{jk} - x_{jk} )}{L}\right)}{\sin \left(\frac{\pi ( t_{ij} -x_{ij} )}{L}\right) \sin \left(\frac{\pi ( t_{ik} -x_{ik} )}{L}\right)}\,.
	\end{aligned}
\end{equation}
The analytic continuation to the real time $t$ has also been considered, and the finite-sized cross-ratios $\xi$ and $\bar \xi$ are as follows,
\begin{equation}\label{cross-ratio-L}
	\begin{aligned}
		\xi = \frac{\sin \left(\frac{\pi(t_{12} + x_{12})}{L}\right) \sin \left(\frac{\pi(t_{34} + x_{34})}{L}\right)}{\sin \left(\frac{\pi ( t_{13} + x_{13} )}{L}\right) \sin \left(\frac{\pi ( t_{24} + x_{24} )}{L}\right)} \,, \\
		\bar \xi =\frac{\sin \left(\frac{\pi(t_{12} - x_{12})}{L}\right) \sin \left(\frac{\pi(t_{34} - x_{34})}{L}\right)}{\sin \left(\frac{\pi ( t_{13} - x_{13} )}{L}\right) \sin \left(\frac{\pi ( t_{24} - x_{24} )}{L}\right)}\,.
	\end{aligned}
\end{equation}
It is worth noting here that the correction for this case in \cref{SR-disj-L} is always negative for spacelike subsystems which indicate that the entangled degrees of freedom between the such subsystems in finite-sized systems also decreases upon the introduction of the $\TTbar$ operator.

\subsubsection*{Timelike entanglement} \label{sec:SR-L-disj-T}
We now consider the configuration of two disjoint (purely) timelike subsystems described by $A \equiv [(x,t_1), (x,t_2)]$ and $B \equiv [(x,t_3), (x,t_4)]$ in a finite-sized CFT$_2$ deformed by a $\TTbar$ operator. Unlike the finite temperature case, we observe that the analytic continuation of \cref{SR-disj-L} to timelike subsystems lead to non-vanishing correction given by
\begin{equation} \label{SR-disj-L-T}
	\begin{aligned}
		\delta S_R^{\text{(T)}} = - \frac{\pi^4 c^2 \mu }{9 L^3} \sqrt{\frac{\sin \left(\frac{\pi t_{13}}{L}\right) \sin \left(\frac{\pi t_{24}}{L}\right)}{\sin \left(\frac{\pi t_{12}}{L}\right) \sin \left(\frac{\pi t_{34}}{L}\right)}} \vast(&\frac{t_1 \sin \left(\frac{\pi t_{34}}{L}\right)}{\sin \left(\frac{\pi t_{13}}{L}\right) \sin \left(\frac{\pi t_{14}}{L}\right)}-\frac{t_2 \sin \left(\frac{\pi t_{34}}{L}\right)}{\sin \left(\frac{\pi t_{23}}{L}\right) \sin \left(\frac{\pi t_{24}}{L}\right)} \\
		&+\frac{t_3 \sin \left(\frac{\pi t_{12}}{L}\right)}{\sin \left(\frac{\pi t_{13}}{L}\right) \sin \left(\frac{\pi  t_{23}}{L}\right)} - \frac{t_4 \sin \left(\frac{\pi  t_{12}}{L}\right)}{\sin \left(\frac{\pi t_{14}}{L}\right) \sin \left(\frac{\pi t_{24}}{L}\right)}\vast) \,.
	\end{aligned}
\end{equation}

\subsubsection{Two adjacent subsystems} \label{sec:SR-L-adj}
In this subsection, we now consider two adjacent boosted subsystems $A \equiv [(w_1,\bar w_1),(w_2,\bar w_2)]$ and $B \equiv [(w_2,\bar w_2),(w_3,\bar w_3)]$ in a finite-sized CFT$_2$ with a $\TTbar$ deformation. The correction to the R\'enyi entropy may be obtained through \cref{Snm-adj-beta} where, now, the complex plane coordinates $(z, \bar z)$ is mapped to the spatial cylinder $(w, \bar w)$ through \cref{cyl-map-L}. Solving the integral in \cref{Snm-adj-beta-int} and subsequently taking the replica limit $n,m \to 1$, we may obtain the correction to the reflected entropy in the leading order as follows
\begin{equation}
	\begin{aligned}\label{SR-adj-L}
		\delta S_R = -\frac{\pi^4 c^2 \mu }{9 L ^3} \vast(\frac{t_{12} \sin \left(\frac{2 \pi t_{12}}{L}\right)}{\cos \left(\frac{2 \pi t_{12}}{L }\right)-\cos \left(\frac{2 \pi x_{12}}{L }\right)}&+\frac{t_{23}\sin \left(\frac{2 \pi t_{23}}{L }\right)}{\cos \left(\frac{2 \pi t_{23}}{L }\right)-\cos \left(\frac{2 \pi x_{23}}{L }\right)}\\
		&-\frac{t_{13} \sin \left(\frac{2 \pi t_{13}}{L}\right)}{\cos \left(\frac{2 \pi t_{13}}{L}\right)-\cos \left(\frac{2 \pi x_{13}}{L }\right)}\vast)\,,
	\end{aligned}
\end{equation}
where we have analytically continued to the real time $t$. Again, the above correction is always negative for spacelike subsystems, indicative of the reduction in entangled degrees of freedom due to the $\TTbar$ operator insertion.

\subsubsection*{Timelike entanglement} \label{sec:SR-L-adj-T}
We now consider the analytic continuation of the above result to the case for (purely) timelike subsystems described by $A \equiv [(x,t_1), (x,t_2)]$ and $B \equiv [(x,t_2), (x,t_3)]$ in a CFT$_2$ deformed by a $\TTbar$ operator defined on a spatial cylinder. The correction to the reflected entropy for such case is non-vanishing and is given by
\begin{equation} \label{SR-adj-L-T}
	\begin{aligned}
		\delta S_R^{\text{(T)}} = - \frac{\pi^4 c^2 \mu }{9 L^3} \left( t_{12} \cot \left(\frac{\pi t_{12}}{L}\right) + t_{23} \cot \left(\frac{\pi t_{23}}{L}\right) - t_{13} \cot \left(\frac{\pi t_{13}}{L}\right)\right)\,.
	\end{aligned}
\end{equation}

\subsubsection{A single subsystem} \label{sec:SR-L-sing}
For the case of a single subsystem in a finite-sized CFT$_2$ perturbed by a $\TTbar$ operator, we consider the subsystem to be described by a boosted spacelike interval $A \equiv [(w_1,\bar w_1),(w_2,\bar w_2)]$. Note that unlike the case for the CFT at a finite temperature, it is not required to introduce the auxiliary intervals in this case, as the branch cuts are finite along the spatial circle. The correction to the R\'enyi reflected entropy is obtained by solving \cref{Snm-ttbar-def} with appropriate two-point twist correlator. Subsequently the correction to the reflected entropy for this configuration is then obtained in the replica limit $n,m \to 1 $ to be
\begin{equation}\label{SR-sing-L}
	\delta S_R = - \frac{\pi^4 c^2 \mu}{9 L^3 }\frac{t_{12} \sin \left( \frac{2 \pi t_{12}}{L} \right)}{\cos \left( \frac{2 \pi t_{12}}{L} \right) - \cos \left( \frac{2 \pi x_{12}}{L} \right)} \,.
\end{equation}
where analytic continuation to real time $t$ has been implemented. Note that, similar to earlier cases, the above correction to the reflected entropy due to the introduction of the $\TTbar$ operator is always negative for spacelike subsystem $A$.

\subsubsection*{Timelike entanglement} \label{sec:SR-L-sing-T}
We now proceed to the case of a single timelike subsystem $A \equiv [(0,0), (0, t)]$ in a finite-sized CFT$_2$ deformed by a $\TTbar$ operator. Note that the timelike subsystem is aligned parallel to the axis of the spatial cylinder. Consequently, as earlier, we encounter infinite branch cuts for this configuration, requiring the introduction of two auxiliary subsystems $B_1 \equiv [(-X,-T) , (0,0)]$ and $B_2 \equiv [(0, t), (X,T)]$. Unlike the finite temperature case, the bipartite limit $B_1 \cup B_2 \equiv B \to A^c$ for the spatial cylinder is implemented by stretching the auxiliary intervals to timelike infinity $T \to \infty$.

The correction to the R\'enyi reflected entropy for this single subsystem sandwiched between two auxiliary subsystems can be obtained through \cref{Snm-sing-beta}. The corresponding four-point twist correlator have the same form as in \cref{four-point-single}, except that the complex plane coordinates $(z, \bar z)$ are now obtained through the map \eqref{cyl-map-L} and $\eta, \bar \eta$ are replaced with the finite-size cross-ratios $\xi, \bar \xi$ given in \cref{cross-ratio-L}. Following arguments similar to \cite{Basu:2022nds, Calabrese:2014yza}, it may be checked that the non-universal functions $\mathcal{F}_{mn}, \bar{\mathcal{F}}_{mn}$ follow similar behaviour as in \cref{non-universal-F} in the limits $\xi, \bar \xi \to 0$ and $\xi, \bar \xi \to 1$. This ultimately leads to the following correction to the reflected entropy in the replica limit $n,m \to 1$, 
\begin{equation} \label{SR-sing-L-T}
	\delta S_R = -\frac{2 \pi^4 c^2 \mu t}{9 L^3} \left[ i - \cot \left(\frac{\pi t}{L}\right) \right] - e^{\frac{2 \pi t}{L}} \frac{f'\left(e^{\frac{2 \pi t}{L}}\right)}{f\left(e^{\frac{2 \pi t}{L}}\right)} - e^{-\frac{2 \pi t}{L}} \frac{\bar {f}'\left(e^{-\frac{2 \pi t}{L}}\right)}{\bar f \left(e^{-\frac{2 \pi t}{L}}\right)}\,,
\end{equation}
where the bipartite limit $T \to \infty$ has been implemented and the non-universal functions $f, \bar f$ given in \cref{non-universal-f} have been introduced. Note that, unlike previous cases, here we also observe correction to the imaginary part of the reflected entropy.

\section{Entanglement wedge cross section in cut-off AdS geometries} \label{sec:EW}
In this section, we compute the EWCS for various bipartite states in the cut-off AdS geometries dual to $\TTbar$ deformed CFT$_2$s and find agreement with the reflected entropy computed from conformal perturbation theory in the limit of a small deformation parameter.
\subsection{Cut-off BTZ black holes} \label{sec:EW-beta}
According to the prescription in \cite{McGough:2016lol}, a thermal CFT$_2$ with $\TTbar$ deformation is dual to the BTZ black hole with the line element
\begin{align}
	\text{d}s^2=-\frac{r^2-r_h^2}{R^2}\text{d}t^2+\frac{R^2}{r^2-r_h^2}\text{d}r^2+\frac{r^2}{R^2}\text{d}\tilde{x}^2\,,
\end{align}
with the finite radial cut-off $r_c=\sqrt{\frac{6R^4}{\pi c \mu}}$. In the above line element, the black hole horizon is located at $r=r_h$ and the time direction is compactified as $t\sim t+i\beta$. It is well known that the black hole has the same temperature as the dual field theory,
\begin{align}
	\beta=\frac{2\pi R^2}{r_h}\,.\label{BTZ-temperature}
\end{align}
In the holographic correspondence described in \cite{McGough:2016lol}, the dual thermal field theory is located at the asymptotic boundary of the spacetime at $r=r_c$ and the metric on this conformal boundary reads
\begin{align}
	\text{d}s^2=-\text{d}t^2+\frac{r_c^2}{r_c^2-r_h^2}\text{d}\tilde{x}^2\equiv-\text{d}t^2+\text{d}x^2\,,
\end{align}
where we have defined the conformal coordinate,
\begin{align}
	x=\tilde{x}\left(1-\frac{r_h^2}{r_c^2}\right)^{-1/2}\,,\label{rescaled-coordinate}
\end{align} 
hence the CFT$_2$ lives on a temporally compactified cylinder described by the coordinates $(t,x)$. 

In the following, we will compute the minimal cross section of the entanglement wedge for various bipartite states involving two disjoint, two adjacent and a single subsystem in the $\TTbar$ deformed thermal CFT$_2$, utilizing the embedding coordinate techniques described in \cref{sec:EW-review}. To simplify latter calculations, we set $R=1$ and introduce a new holographic coordinate $u=\frac{1}{r}$, with $u_c=\frac{1}{r_c}$ and $u_h=\frac{1}{r_h}$. The embedding coordinate transformations which map the BTZ metric to that of $\mathbb{R}^{2,2}$ are given by,
\begin{align}
	&X_0(u,t,\tilde{x})=\sqrt{\frac{u_h^2}{u^2}-1}\,\sinh\left(\frac{t}{u_h}\right)\,,\notag\\
	&X_1(u,t,\tilde{x})=\frac{u_h}{u}\,\cosh\left(\frac{\tilde{x}}{u_h}\right)\,,\notag\\
	&X_2(u,t,\tilde{x})=\sqrt{\frac{u_h^2}{u^2}-1}\,\cosh\left(\frac{t}{u_h}\right)\,,\notag\\
	&X_3(u,t,\tilde{x})=\frac{u_h}{u}\,\sinh\left(\frac{\tilde{x}}{u_h}\right)\,.\label{embedding-BTZ}
\end{align}
\subsubsection{Two disjoint subsystems} \label{sec:EW-beta-disj}
\begin{figure}[h!]
	\centering
	\includegraphics[scale=0.7]{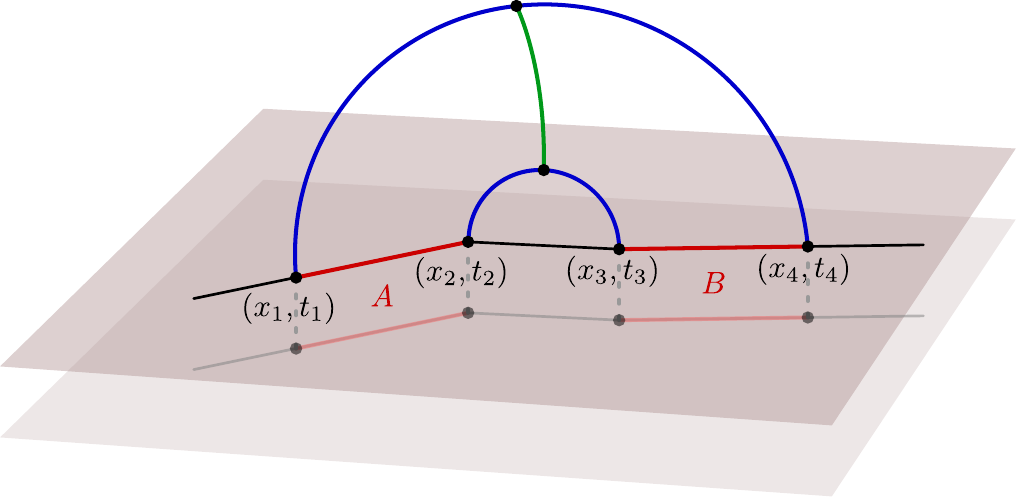}
	\caption{Schematics of two Lorentz boosted disjoint subsystems $A$ and $B$ in the cut-off BTZ geometry. The two shaded planes represent the original and the shifted asymptotic boundary, and the corresponding scaling of the subsystems. Green curve is the EWCS.}
	\label{fig:disj-ttbar}
\end{figure}

We begin with two generic boosted disjoint subsystems $A=[(x_1,t_1),(x_2,t_2)]$ and $B=[(x_3,t_3),(x_4,t_4)]$ as depicted in \cref{fig:disj-ttbar}. Utilizing \cref{embedding-BTZ,EW-embedding-disj}, the EWCS corresponding to the reduced density matrix $\rho_{AB}$ is then given by \cref{EW-embedding-disj}, with
\begin{align}
	\zeta_{ij}&=\left(\frac{u_h^2}{u_c^2}-1\right) \cosh \left(\frac{t_{ij}}{u_h}\right)-\frac{u_h^2}{u_c^2} \cosh \left(\frac{\sqrt{u_h^2-u_c^2} }{u_h^2}\,|x_{ij}|\right)\,.
\end{align}
To compare with the field theoretic result, we consider the limit of small deformation parameter $\mu$ which corresponds to a large cut-off radius $r_c$ (or, small $u_c$). Furthermore, as the bulk BTZ black hole is dual to the finite temperature deformed CFT$_2$ in the limit of high temperatures, we further consider the limit $\beta\ll |x_{ij}|,t_{ij}$. Expanding the expression of the EWCS for small $u_c$ and $u_h\ll |x_{ij}|,t_{ij}$, we may obtain
\begin{align}
	E_W(A:B)=&\frac{1}{4G_N}\cosh^{-1}\left[\frac{1+\sqrt{\displaystyle\eta\,\bar{\eta}}}{\sqrt{(1-\eta)(1-\bar{\eta})}}\right]-\frac{u_c^2\sqrt{\eta\,\bar{\eta}}}{16 G_N u_h^3 \left(\sqrt{\eta}+\sqrt{\bar{\eta}}\right)}\Bigg[\frac{x_{12}\sinh\left(\frac{x_{12}}{u_h}\right)}{\cosh\left(\frac{t_{12}}{u_h}\right)-\cosh\left(\frac{x_{12}}{u_h}\right)}\notag\\
	&+\frac{x_{34}\sinh\left(\frac{x_{34}}{u_h}\right)}{\cosh\left(\frac{t_{34}}{u_h}\right)-\cosh\left(\frac{x_{34}}{u_h}\right)}-\frac{x_{23}\sinh\left(\frac{x_{23}}{u_h}\right)}{\cosh\left(\frac{t_{23}}{u_h}\right)-\cosh\left(\frac{x_{23}}{u_h}\right)}-\frac{x_{14}\sinh\left(\frac{x_{14}}{u_h}\right)}{\cosh\left(\frac{t_{14}}{u_h}\right)-\cosh\left(\frac{x_{14}}{u_h}\right)}\Bigg]\notag\\
	&-\frac{u_c^2}{16 G_N u_h^3 \left(\sqrt{\eta}+\sqrt{\bar{\eta}}\right)}\Bigg[\frac{x_{13}\sinh\left(\frac{x_{13}}{u_h}\right)}{\cosh\left(\frac{t_{13}}{u_h}\right)-\cosh\left(\frac{x_{13}}{u_h}\right)}+\frac{x_{24}\sinh\left(\frac{x_{24}}{u_h}\right)}{\cosh\left(\frac{t_{24}}{u_h}\right)-\cosh\left(\frac{x_{24}}{u_h}\right)}\notag\\
	&\qquad\qquad\qquad\qquad\qquad-\frac{x_{23}\sinh\left(\frac{x_{23}}{u_h}\right)}{\cosh\left(\frac{t_{23}}{u_h}\right)-\cosh\left(\frac{x_{23}}{u_h}\right)}-\frac{x_{14}\sinh\left(\frac{x_{14}}{u_h}\right)}{\cosh\left(\frac{t_{14}}{u_h}\right)-\cosh\left(\frac{x_{14}}{u_h}\right)}\Bigg]\,,\label{EW-disj-BTZ}
\end{align}
where $\eta$ and $\bar{\eta}$ are the finite temperature cross-ratios given in \cref{cross-ratio-generic}. The first term in the above expression is nothing but the EWCS corresponding to the two disjoint subsystems in the undeformed CFT$_2$ located at the asymptotic boundary $r\to\infty$ of the bulk BTZ black hole geometry \cite{Dutta:2019gen}. Utilizing the holographic dictionary in \cref{rc-generic,BTZ-temperature} along with the usual Brown-Henneaux relation is AdS$_3$/CFT$_2$ \cite{Brown:1986nw}, the rest of the terms is easily seen to match with the corresponding field theoretic computations for the corrections to the reflected entropy given in \cref{SR-disj-beta}.
\subsubsection*{Timelike entanglement} \label{sec:EW-beta-disj-T}
We now consider the correlation between two purely timelike subsystems $A=[(x,t_1),(x,t_2)]$ and $B=[(x,t_3),(x,t_4)]$, in the spirit of the timelike entanglement introduced in \cite{Doi:2022iyj,Doi:2023zaf}. While an explicit geometric construction of a timelike entanglement wedge remains absent, we adopt a straightforward approach by accepting its inherent properties and proceed with a rudimentary analytic continuation of our result for the EWCS in \cref{EW-disj-BTZ}. It is easy to verify that the correction to the EWCS due to the deformation vanishes identically,
\begin{align}
	E_W(A:B)=&\frac{1}{4G_N}\cosh^{-1}\left[1+2\frac{\sinh \left(\frac{t_{12}}{2u_h}\right) \sinh \left(\frac{t_{34}}{2u_h}\right)}{\sinh \left(\frac{ t_{23}}{2u_h}\right) \sinh \left(\frac{ t_{14}}{2u_h}\right)}\right]\,.\label{disj-timelike-BTZ}
\end{align}
Upon utilizing the Brown-Henneaux formula \cite{Brown:1986nw} and the holographic dictionary in \cref{BTZ-temperature}, the above expression for the EWCS matches identically with the field theoretic result for the reflected entropy in \cref{SR-disj-beta-T}.
Interestingly the corrections due to $\TTbar$ deformation is absent in the above expression as the angular separations of the timelike intervals are unaffected while pushing the holographic screen inwards \cite{Jiang:2023ffu,Basu:2023aqz}. Furthermore, we notice a vanishing imaginary contribution to the EWCS indicating no timelike curve segment in the bulk construction. This ought to shed some light into the geometry of the corresponding entanglement wedge.
\subsubsection{Two adjacent subsystems} \label{sec:EW-beta-adj}
\begin{figure}[h!]
	\centering
	\includegraphics[scale=0.7]{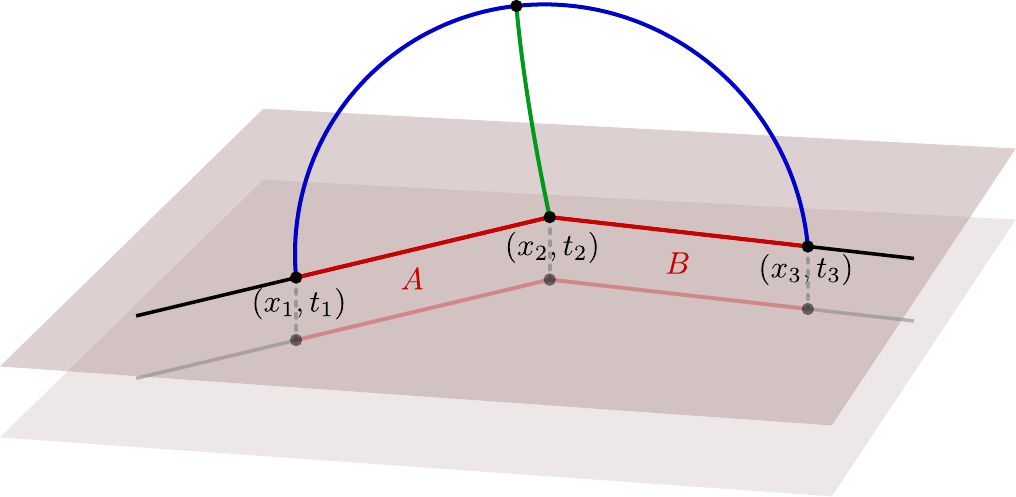}
	\caption{Schematics of two Lorentz boosted adjacent subsystems $A$ and $B$ in the cut-off BTZ geometry.}
	\label{fig:adj-ttbar}
\end{figure}

Next, we consider the case of two adjacent boosted subsystems $A=[(x_1,t_1),(x_2,t_2)]$ and $B=[(x_2,t_2),(x_3,t_3)]$ in the dual thermal CFT$_2$ with $\TTbar$ deformation as depicted in \cref{fig:adj-ttbar}. Utilizing \cref{EW-embedding-adj} and the embedding coordinates given in \cref{embedding-BTZ}, the EWCS between the two adjacent subsystems may be obtained as follows
\begin{align}
	E_W(A:B)=\frac{1}{4G_N}\cosh^{-1}\left(\sqrt{\displaystyle\frac{2\left[\alpha^2\cosh \left(\frac{t_{12}}{u_h}\right)-\frac{u_h^2}{u_c^2} \cosh \left(\frac{\alpha |x_{12}|}{u_h}\right)\right]\left[\alpha^2\cosh \left(\frac{t_{23}}{u_h}\right)-\frac{u_h^2}{u_c^2} \cosh \left(\frac{\alpha |x_{23}|}{u_h}\right)\right]}{\left[\alpha^2\cosh \left(\frac{t_{13}}{u_h}\right)-\frac{u_h^2}{u_c^2} \cosh \left(\frac{\alpha |x_{13}|}{u_h}\right)\right]}}\right),\label{EW-adj-BTZ-complete}
\end{align}
where we have defined $\alpha^2=\frac{u_h^2}{u_c^2}-1$. Once again, we consider the limit of small deformation parameter (small $u_c$) at a large temperature and expand the above expression to obtain
\begin{align}
	E_W(A:B)=\frac{1}{8G_N}&\log\left(\frac{8u_h^2\left[\cosh\left(\frac{t_{12}}{u_h}\right)-\cosh\left(\frac{x_{12}}{u_h}\right)\right]\left[\cosh\left(\frac{t_{23}}{u_h}\right)-\cosh\left(\frac{x_{23}}{u_h}\right)\right]}{u_c^2\left[\cosh\left(\frac{t_{13}}{u_h}\right)-\cosh\left(\frac{x_{13}}{u_h}\right)\right]}\right)\notag\\
	&-\frac{u_c^2}{16G_N u_h^3}\Bigg[\frac{x_{12}\sinh\left(\frac{x_{12}}{u_h}\right)}{\cosh\left(\frac{t_{12}}{u_h}\right)-\sinh\left(\frac{x_{12}}{u_h}\right)}+\frac{x_{23}\sinh\left(\frac{x_{23}}{u_h}\right)}{\cosh\left(\frac{t_{23}}{u_h}\right)-\sinh\left(\frac{x_{23}}{u_h}\right)}\notag\\
	&\qquad\qquad\qquad\qquad\qquad\qquad\qquad\qquad-\frac{x_{13}\sinh\left(\frac{x_{13}}{u_h}\right)}{\cosh\left(\frac{t_{13}}{u_h}\right)-\sinh\left(\frac{x_{13}}{u_h}\right)}\Bigg]\label{EW-adj-BTZ}
\end{align}
Utilizing the holographic dictionary in \cref{rc-generic} and the expression for the temperature of the black hole in \cref{BTZ-temperature}, the first term in the above expression may be identified as the EWCS between two adjacent subsystems in the original undeformed CFT$_2$ with UV cut-off $\epsilon_c$. On the other hand, the sub-leading terms proportional to $u_c^2$ match identically with the leading order corrections to the corresponding reflected entropy obtained from conformal perturbation theory in \cref{SR-adj-beta}, upon utilizing the Brown-Henneaux relation in AdS$_3$/CFT$_2$.
\subsubsection*{Timelike entanglement} \label{sec:EW-beta-adj-T}
We consider the case of two purely timelike adjacent subsystems $A=[(x,t_1),(x,t_2)]$ and $B=[(x,t_2),(x,t_3)]$ in the thermal CFT$_2$ with $\TTbar$ deformation. Once again, despite the absence of an explicit geometric construction of the corresponding entanglement wedge, we accept its existence prima facie and adopt an analytic continuation of the result \cref{EW-adj-BTZ} for two boosted adjacent intervals, to obtain
\begin{align}
	E_W(A:B) = \frac{1}{4G_N}\left(\log\Bigg[\frac{2 u_c}{u_h}\,\frac{\sinh{\big(\frac{ t_{12}}{2u_h}\big)}\sinh{\big(\frac{ t_{23}}{2u_h}\big)}}{\sinh({\frac{t_{13}}{2u_h} })}\Bigg]+i\pi\right)\,.
\end{align}
As in the case of two disjoint subsystems, we observe that the corrections due to the $\TTbar$ deformation vanishes identically owing to the geometric orientation of the subsystems along the thermal cylinder \cite{Jiang:2023ffu,Basu:2023aqz}. Utilizing the holographic dictionary in \cref{rc-generic,BTZ-temperature} and the Brown-Henneaux relation \cite{Brown:1986nw}, the above expression is easily seen to conform to the result in \cref{SR-adj-beta-timelike}.

\subsubsection{A single subsystem} \label{sec:EW-beta-sing}
\begin{figure}[h!]
	\centering
	\includegraphics[scale=0.7]{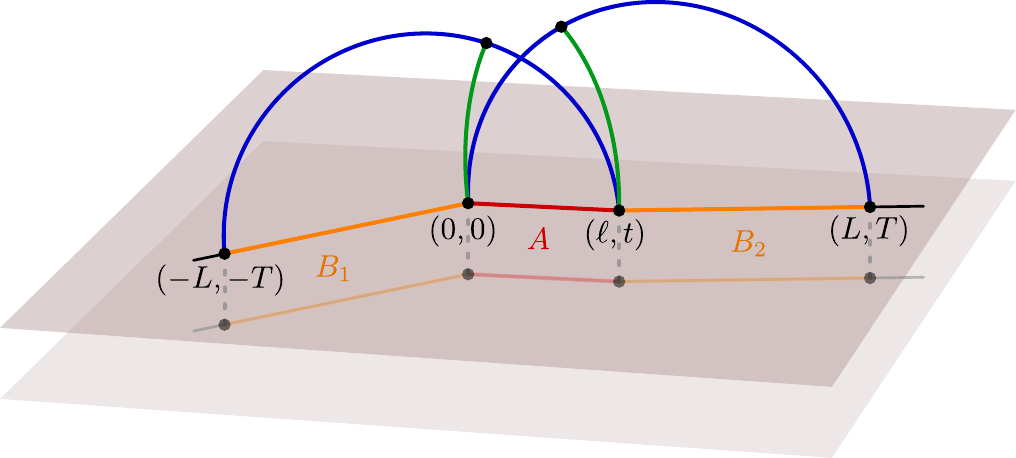}
	\caption{Schematics of a Lorentz boosted subsystem $A$ in the cut-off BTZ geometry. $B_1$ and $B_2$ are auxiliary intervals. Union of the green curves gives the upper bound to the EWCS for this configuration.}
	\label{fig:sing-ttbar}
\end{figure}

In this subsection, we consider a boosted spacelike subsystem $A=[(0,0),(\ell,t)]$ in the $\TTbar$ deformed thermal CFT$_2$. As described earlier, this requires the introduction of two large auxiliary subsystems $B_1=[(-L,-T),(0,0)]$ and $B_2=[(\ell,t),(L,T)]$ sandwiching the single subsystem in question which is depicted in \cref{fig:sing-ttbar}. Subsequently, we construct the bulk co-dimension one entanglement wedge dual to the bipartite density matrix $\rho_{A\cup B}$ with $B\equiv B_1\cup B_2$. We may compute the upper bound to the EWCS between $A$ and $B$ utilizing the following expression \cite{KumarBasak:2020eia,Basu:2021awn,Basu:2022nds}
\begin{align}
	\tilde{E}_W(A:B)=E_W(A:B_1)+E_W(A:B_2)\,.\label{EW-upper-bound}
\end{align}
The upper bound to the EWCS corresponding to the single subsystem in question may then be obtained by taking the bipartite limit $B\to A^c$ ($L\to\infty$). Noting the fact that $A,B_i$ are adjacent to each other, we may utilize the result \cref{EW-adj-BTZ} to compute the individual EWCS $E_W(A:B_i)$ and obtain
\begin{align}
	\tilde{E}_W(A:B)&=\frac{1}{4G_N}\log\left(\frac{8u_h^2}{u_c^2}\left[\cosh \left(\frac{\ell}{u_h}\right)-\cosh \left(\frac{t}{u_h}\right)\right] \sqrt{\displaystyle\frac{\cosh \left(\frac{\ell-L}{u_h}\right)-\cosh \left(\frac{t-T}{u_h}\right)}{\cosh \left(\frac{\ell+L}{u_h}\right)-\cosh \left(\frac{t+T}{u_h}\right)}}\right)\notag\\
	&-\frac{u_c^2}{16G_N u_h^3}\Bigg[\frac{2 \ell \sinh \left(\frac{\ell}{u_h}\right)}{\cosh \left(\frac{t}{u_h}\right)-\cosh \left(\frac{\ell}{u_h}\right)}+\frac{(\ell+L) \sinh \left(\frac{\ell+L}{u_h}\right)}{\cosh \left(\frac{t+T}{u_h}\right)-\cosh \left(\frac{\ell+L}{u_h}\right)}\notag\\
	&\qquad\qquad\qquad\qquad\qquad\qquad\qquad\qquad-\frac{(\ell-L) \sinh \left(\frac{\ell-L}{u_h}\right)}{\cosh \left(\frac{t-T}{u_h}\right)-\cosh \left(\frac{\ell-L}{u_h}\right)}\Bigg]\,.
\end{align}
The upper bound to the EWCS corresponding to the single boosted subsystem in question may now be obtained by taking the bipartite limit $L\to \infty$ as follows
\begin{align}
	\tilde{E}_W(A:A^c)=&\frac{1}{4G_N}\left[\log\left(\frac{2u_h^2}{u_c^2}\left[\cosh \left(\frac{\ell}{u_h}\right)-\cosh \left(\frac{t}{u_h}\right)\right]\right)-\frac{\ell}{ u_h}\right]+\frac{1}{2G_N}\log 2\notag\\
	&-\frac{ u_c^2\,\ell }{8 G_N u_h^3}\left[1+\frac{\sinh \left(\frac{\ell}{u_h}\right)}{\cosh \left(\frac{t}{u_h}\right)-\cosh \left(\frac{\ell}{u_h}\right)}\right]\label{EW-sing-final}
\end{align}
As earlier, the first term in the above expression is just the EWCS corresponding to the single subsystem in the undeformed CFT$_2$, whereas the term proportional to $u_c^2$ is the leading order correction due to the $\TTbar$ deformation. The constant term may be attributed to the Markov gap \cite{Hayden:2021gno} owing to its geometrical interpretation in terms of the number of non-trivial boundaries of the EWCS. Utilizing the holographic dictionary in \cref{rc-generic,BTZ-temperature} and the Brown-Henneaux relation in AdS$_3$/CFT$_2$, the sub-leading term may be seen to match exactly\footnote{\label{footnote-single-BTZ}The non-universal functions $f$ and $\bar{f}$ are expected to be sub-leading in the large central charge limit and hence are undetected in the holographic computations. Furthermore, the constant mismatch of $\log 2/(2G_N)$ may be attributed to the Markov gap \cite{Hayden:2021gno} pertaining to our holographic construction for the upper bound in terms of the EWCS between two adjacent intervals.} with the corresponding leading order correction to the reflected entropy obtained from conformal perturbation theory in \cref{SR-sing-beta}. Interestingly, the above expression for the EWCS in the dual cut-off geometry maybe recast into the following instructive form
\begin{align}
	\tilde{E}_W(A:A^c)=S_A-S_A^{\text{Th}}+\frac{1}{2G_N}\log 2\,,
\end{align}
where $S_A$ is the entanglement entropy of the single subsystem under consideration and $S_A^{\text{Th}}$ is the corresponding thermal entropy
\begin{align}
	S_A^{\text{Th}}=\frac{\ell}{4G_N u_h}\left(1-\frac{u_c^2}{2\,u_h^2}\right)=\frac{\pi c\,\ell}{3\beta}\left(1-\mu\frac{\pi^3c}{3\,\beta^2}\right)\,.\label{Thermal-entropy}
\end{align}
Hence, we see that the thermal entropy gets non-trivial corrections due to the $\TTbar$ deformation. This  may be understood from the re-scaling of the spatial coordinate given in \cref{rescaled-coordinate} in the limit of small deformation parameter $\mu$. It is well known that the thermal contribution to the entanglement entropy for a single subsystem arises as the corresponding HRT surface wraps the black hole horizon \cite{Hubeny:2007xt}. The correction to the thermal entropy due to the $\TTbar$ deformation may then be interpreted from the fact that as the holographic screen is pushed inside the bulk, the wrapping of the extremal surface around the black hole horizon decreases.
\subsubsection*{Timelike entanglement} \label{sec:EW-beta-sing-T}
Finally, we consider a purely timelike subsystem $A=[(x_0,0),(x_0,t)]$ in the thermal $\TTbar$ deformed CFT$_2$. In contrast to the spacelike subsystem considered earlier, in this case, we do not require any auxiliary subsystems to remove the pathology of tracing out an infinite branch cut. Hence, the EWCS corresponding to the single subsystem in question reduces to the minimal surface homologous to the subsystem. The length of such minimal surface has already been computed in \cite{Doi:2022iyj, Doi:2023zaf}. The EWCS in this case is therefore given by
\begin{align}
	E_W(A:A^c)=\frac{1}{2G_N}\left(\log\left[\frac{2u_h}{u_c}\sinh\left(\frac{t}{u_h}\right)\right]+\frac{i\pi}{2}\right)\,,
\end{align} 
which matches identically with half of the reflected entropy for the present configuration \cite{Afrasiar:2022wzn}. Interestingly, the above result may be obtained from an analytic continuation of the first term in \cref{EW-sing-final} which serves as a strong consistency check of our holographic construction.
\subsection{Cut-off global AdS} \label{sec:EW-L}
In this section, we consider the finite cut-off global AdS geometry dual to a $\TTbar$ deformed CFT$_2$ with a finite size. The line element for the global AdS$_3$ geometry reads
\begin{align}
	\text{d}s^2=R^2\left(-\cosh^2\rho\, \text{d}\tau^2+\text{d}\rho^2+\sinh^2\rho \,\text{d}\phi^2\right)\label{global-metric}
\end{align}
where the spatial direction $\phi$ is compactified with period $2\pi$, $\phi\sim\phi+2\pi$. According to the prescription in \cite{McGough:2016lol}, the dual finite-sized CFT$_2$ is located at the radial cut-off $\rho_c$ where
\begin{align}
	\rho_c=\cosh^{-1}\left(\sqrt{\frac{3L^2}{2\mu\pi^3 c}}\right)\,,\label{rho-c-global}
\end{align}
where $L$ is the circumference of the boundary circle. The UV cut-off of the CFT$_2$ may then be related to the bulk radial cut-off utilizing the relation in \cref{rc-generic} as follows
\begin{align}
	\cosh \rho_c=\frac{L}{2\pi\epsilon_c}\,.\label{global-cut-off}
\end{align}
Therefore, the line element at the cut-off boundary is given by
\begin{align}
	\text{d}s^2=-\coth^2\rho_c\,\text{d}\tau^2+\text{d}\phi^2\equiv -\text{d}\theta^2+\text{d}\phi^2\,,\label{Finite-size-bdy-metric}
\end{align}
where we have defined the conformal time coordinate 
\begin{align}
	\theta=\frac{\tau}{\tanh\rho_c\,}\,,
\end{align}
such that the CFT$_2$ is defined on a cylinder compactified along the spatial direction with circumference $L$. The CFT$_2$ may be described by the coordinates $(t,x)$, which are related to the bulk coordinates as follows, $\theta=\frac{2\pi t}{L}\,,\,\phi=\frac{2\pi x}{L}$.

We may embed the geometry described by the metric \cref{global-metric} in $\mathbb{R}^{2,2}$ by utilizing the following coordinate transformations 
\begin{align}
	&X_0(\theta,\phi,\rho)=R \cosh \rho\sin\theta~~,~~X_1(\theta,\phi,\rho)=R \cosh \rho\cos\theta\,,\notag\\
	&X_2(\theta,\phi,\rho)=R \sinh \rho\cos\phi~~,~~X_3(\theta,\phi,\rho)= R \sinh \rho\sin\phi\,.\label{embedding-Global}
\end{align}
In these embedding coordinates, the distance between two arbitrary points $X_i:(\theta_i,\phi_i,\rho_c)$ and $X_j:(\theta_j,\phi_j,\rho_c)$ on the cut-off boundary may be computed straightforwardly as follows
\begin{align}
	\zeta_{ij}=-X_i\cdot X_j&=R^2\left[\sinh^2\rho_c\,\cos\tau_{12}-\cosh^2\rho_c\,\cos\phi_{12}\right]\notag\\
	&=R^2\left[\sinh^2\rho_c\,\cos\left(\frac{2\pi\, x_{12}}{L}\right)-\cosh^2\rho_c\,\cos\left(\frac{2\pi\,t_{12}}{L}\tanh\rho_c\right)\right]\,.\label{coordinate-distance-global}
\end{align}
In the following, we will compute the minimal cross section of the entanglement wedge for various bipartite states involving two disjoint, two adjacent and a single subsystem in the $\TTbar$ deformed finite-sized CFT$_2$, utilizing the embedding coordinates in \cref{embedding-Global} and the distance formula \cref{coordinate-distance-global}. For brevity of notations, in the following, we will set the AdS radius to unity, $R=1$.
\subsubsection{Two disjoint subsystems} \label{sec:EW-L-disj}
We begin with the case of two disjoint subsystems $A=[(x_1,t_1),(x_2,t_2)]$ and $B=[(x_3,t_3),(x_4,t_4)]$ in a $\TTbar$ deformed finite-sized CFT$_2$. Utilizing the embedding coordinates in \cref{embedding-Global}, we may compute the EWCS corresponding to the mixed state under consideration from \cref{EW-embedding-disj} as follows
\begin{align}
	E_W(A:B)=\frac{1}{4G_N}\cosh^{-1}\Bigg[&\sqrt{\displaystyle \frac{\left(\tanh^2\rho_c\,\cos\tau_{31}-\cos\phi_{31}\right)\left(\tanh^2\rho_c\,\cos\tau_{42}-\cos\phi_{42}\right)}{\left(\tanh^2\rho_c\,\cos\tau_{32}-\cos\phi_{32}\right)\left(\tanh^2\rho_c\,\cos\tau_{41}-\cos\phi_{41}\right)}}
	\notag\\
	&+\sqrt{\displaystyle \frac{\left(\tanh^2\rho_c\,\cos\tau_{21}-\cos\phi_{21}\right)\left(\tanh^2\rho_c\,\cos\tau_{43}-\cos\phi_{43}\right)}{\left(\tanh^2\rho_c\,\cos\tau_{32}-\cos\phi_{32}\right)\left(\tanh^2\rho_c\,\cos\tau_{41}-\cos\phi_{41}\right)}}\Bigg]\label{EW-disj-Global-full}
\end{align}
In order to compare with field theoretic results for the reflected entropy, we consider the limit of a small deformation parameter $\mu$, or large cut-off radius $\rho_c$. Expanding \cref{EW-disj-Global-full} for large $\rho_c$, namely for $L\gg |X_{ij}|,t_{ij}$ (cf. \cref{rho-c-global}), we obtain to leading order in the cut-off scale $\chi_c=\frac{1}{\cosh \rho_c}$,
\begin{align}
	E_W(A:B)=&\frac{1}{4G_N}\cosh^{-1}\left[\frac{1+\sqrt{\displaystyle\xi\,\bar{\xi}}}{\sqrt{\displaystyle(1-\xi)(1-\bar{\xi})}}\right]-\frac{\pi \chi_c^2\sqrt{\xi\,\bar{\xi}}}{8 G_N L  \left(\sqrt{\xi}+\sqrt{\bar{\xi}}\right)}\Bigg[\frac{t_{12}\sin\left(\frac{2\pi t_{12}}{L}\right)}{\cos\left(\frac{2\pi t_{12}}{L}\right)-\cos\left(\frac{2\pi x_{12}}{L}\right)}\notag\\
	&+\frac{t_{34}\sin\left(\frac{2\pi t_{34}}{L}\right)}{\cos\left(\frac{2\pi t_{34}}{L}\right)-\cos\left(\frac{2\pi x_{34}}{L}\right)}-\frac{t_{23}\sin\left(\frac{2\pi t_{23}}{L}\right)}{\cos\left(\frac{2\pi t_{23}}{L}\right)-\cos\left(\frac{2\pi x_{23}}{L}\right)}-\frac{t_{14}\sin\left(\frac{2\pi t_{14}}{L}\right)}{\cos\left(\frac{2\pi t_{14}}{L}\right)-\cos\left(\frac{2\pi x_{14}}{L}\right)}\Bigg]\notag\\
	&-\frac{\pi\chi_c^2 }{8 G_N L \left(\sqrt{\xi}+\sqrt{\bar{\xi}}\right)}\Bigg[\frac{t_{13}\sin\left(\frac{2\pi t_{13}}{L}\right)}{\cos\left(\frac{2\pi t_{13}}{L}\right)-\cos\left(\frac{2\pi x_{13}}{L}\right)}+\frac{t_{24}\sin\left(\frac{2\pi t_{24}}{L}\right)}{\cos\left(\frac{2\pi t_{24}}{L}\right)-\cos\left(\frac{2\pi x_{24}}{L}\right)}\notag\\
	&\qquad\qquad\qquad-\frac{t_{23}\sin\left(\frac{2\pi t_{23}}{L}\right)}{\cos\left(\frac{2\pi t_{23}}{L}\right)-\cos\left(\frac{2\pi x_{23}}{L}\right)}-\frac{t_{14}\sin\left(\frac{2\pi t_{14}}{L}\right)}{\cos\left(\frac{2\pi t_{14}}{L}\right)-\cos\left(\frac{2\pi x_{14}}{L}\right)}\Bigg]+\mathcal{O}\left(\chi_c^{3}\right)\,,\label{EW-disj-Global}
\end{align}
where $\xi$ and $\bar{\xi}$ are the cross-ratios corresponding to the two disjoint subsystems in the finite-sized CFT$_2$ given in \cref{finite-size-cross-ratio}.
\begin{align}
	\xi = \frac{\sin\left(\frac{\pi(t_{12} + x_{12})}{L}\right) \sin\left(\frac{\pi(t_{34} + x_{34})}{L}\right)}{\sin \left(\frac{\pi ( t_{13} + x_{13} )}{L}\right) \sin \left(\frac{\pi ( t_{24} + x_{24} )}{L}\right)} \,, \notag\\
	\bar \xi =\frac{\sin \left(\frac{\pi(t_{12} - x_{12})}{L}\right) \sin \left(\frac{\pi(t_{34} - x_{34})}{L}\right)}{\sin \left(\frac{\pi ( t_{13} - x_{13} )}{L}\right) \sin \left(\frac{\pi ( t_{24} - x_{24} )}{L}\right)}\,.\label{finite-size-cross-ratio}
\end{align}
As in the finite temperature case, the first term in the above expression may be identified with the EWCS between the two disjoint subsystems in the absence of the $\TTbar$ deformation. The rest of the terms proportional to $\chi_c^2$ denote the leading order corrections to the EWCS due to the deformation. Utilizing the holographic dictionary in \cref{rho-c-global} and the Brown-Henneaux formula, the leading order corrections may be seen to match exactly with the corresponding field theoretic computations for the corrections to the reflected entropy given in \cref{SR-disj-L}.

\subsubsection*{Timelike entanglement} \label{sec:EW-L-disj-T}
For two purely timelike disjoint intervals $A=[(x,t_1),(x,t_2)]$ and $B=[(x,t_3),(x,t_4)]$, once again, we forego the pursuit of an explicit geometric delineation of the entanglement wedge and instead undertake a naive approach to the bulk dual for the reflected entropy. We proceed by analytically continuing our result for the EWCS for two boosted spacelike intervals given in \cref{EW-disj-Global}, to obtain the following leading order expression for a small deformation parameter,
\begin{align}
	E_W(A:B)=&\frac{1}{4G_N}\cosh^{-1}\left[1+2\frac{\sin \left(\frac{\pi t_{12}}{L}\right) \sin \left(\frac{\pi t_{34}}{L}\right)}{\sin\left(\frac{\pi t_{23}}{L}\right) \sin\left(\frac{\pi t_{14}}{L}\right)}\right]\notag\\
	&-\frac{\pi\chi_c^2 }{8 G_N L}\sqrt{\displaystyle\frac{\sin \left(\frac{\pi t_{12}}{L}\right) \sin \left(\frac{\pi t_{34}}{L}\right)}{\sin\left(\frac{\pi t_{13}}{L}\right) \sin\left(\frac{\pi t_{24}}{L}\right)}}\Bigg[t_{12} \cot \left(\frac{\pi t_{12}}{L}\right)+t_{34} \cot \left(\frac{\pi t_{34}}{L}\right)\notag\\&\qquad\qquad\qquad\qquad\qquad\qquad\quad~-t_{23} \cot \left(\frac{\pi t_{23}}{L}\right)-t_{14} \cot \left(\frac{\pi t_{14}}{L}\right)\Bigg]\,.
\end{align}
In the above expression, the first term concerns the EWCS for two timelike disjoint intervals for a undeformed CFT$_2$ defined on a spatially compactified cylinder, while the rest of the expression corresponds to the contributions due to the $\TTbar$ deformation owing to an inward displacement of the holographic screen. Utilizing the holographic dictionary in \cref{rho-c-global} and the usual Brown-Henneaux formula in the context of AdS$_3$/CFT$_2$, it is straightforward to verify that the above expression is tantamount to half the reflected entropy computed in \cref{SR-disj-L-T}. Note that, similar to the case of the BTZ black hole geometry (cf. \cref{disj-timelike-BTZ}), there is no imaginary contribution to the reflected entropy for two disjoint timelike intervals indicating the absence of a timelike curve segment in the corresponding geometric construction.
\subsubsection{Two adjacent subsystems} \label{sec:EW-L-adj}
Next,we consider two adjacent subsystems $A=[(x_1,t_1),(x_2,t_2)]$ and $B=[(x_2,t_2),(x_3,t_3)]$ in the finite-sized CFT$_2$ with a $\TTbar$ deformation. Utilizing \cref{EW-embedding-adj} and the coordinate transformations in \cref{embedding-Global}, we may obtain the following expression for the EWCS corresponding to the mixed state under consideration,
\begin{align}
	E_W(A:B)=\frac{1}{4G_N}\cosh^{-1}\Bigg[\sqrt{\frac{2\,\chi_c^2\left(\tanh^2\rho_c\,\cos\tau_{21}-\cos\phi_{21}\right)\left(\tanh^2\rho_c\,\cos\tau_{32}-\cos\phi_{32}\right)}{\left(\tanh^2\rho_c\,\cos\tau_{31}-\cos\phi_{31}\right)}}\Bigg]\,,\label{EW-adj-Global-full}
\end{align}
where the boundary coordinates $(x_i,t_i)$ are related to the $(\tau_i,\phi_i)$ coordinates as follows
\begin{align}
	\tau_i=\frac{2\pi\,t_i}{L}\tanh\rho_c~~,~~\phi_i=\frac{2\pi\,x_i}{L}\,.
\end{align}
We may expand \cref{EW-adj-Global-full} for large cut-off radius $\rho_c$ (corresponding to the small deformation parameter $\mu$) to obtain
\begin{align}
	E_W(A:B)=&\frac{1}{8G_N}\log\left[\left(\frac{2L}{\pi\,\epsilon_c}\right)^2\frac{\left[\cos\left(\frac{2\pi x_{21}}{L}\right)-\cos\left(\frac{2\pi t_{21}}{L}\right)\right]\left[\cos\left(\frac{2\pi x_{32}}{L}\right)-\cos\left(\frac{2\pi t_{32}}{L}\right)\right]}{2\left[\cos\left(\frac{2\pi x_{31}}{L}\right)-\cos\left(\frac{2\pi t_{31}}{L}\right)\right]}\right]\notag\\
	&-\frac{\pi \chi_c^2}{8G_N L}\Bigg[\frac{t_{21}\sin\left(\frac{2\pi t_{21}}{L}\right)}{\cos\left(\frac{2\pi t_{21}}{L}\right)-\cos\left(\frac{2\pi x_{21}}{L}\right)}+\frac{t_{32}\sin\left(\frac{2\pi t_{32}}{L}\right)}{\cos\left(\frac{2\pi t_{32}}{L}\right)-\cos\left(\frac{2\pi x_{32}}{L}\right)}\notag\\
	&\qquad\qquad\qquad\qquad\qquad\qquad\qquad\qquad\,\,-\frac{t_{31}\sin\left(\frac{2\pi t_{31}}{L}\right)}{\cos\left(\frac{2\pi t_{31}}{L}\right)-\cos\left(\frac{2\pi x_{31}}{L}\right)}\Bigg]+\mathcal{O}\left(\chi_c^{3}\right)\,,\label{EW-adj-Global}
\end{align}
where in the first term we have utilized the relation \cref{global-cut-off} to convert the bulk cut-off scale $\chi_c$ to the UV cut-off $\epsilon_c$ in the dual field theory. As earlier, we may identify the first term on the right hand side of the above expression as the EWCS between the two adjacent subsystems in an undeformed CFT$_2$. Furthermore, utilizing \cref{rho-c-global} and the usual Brown-Henneaux relation in AdS$_3$/CFT$_2$, the rest of the terms proportional to $\chi_c^{-2}$ match identically to the leading order corrections to the reflected entropy due to the $\TTbar$ deformation, given in \cref{SR-adj-L}.

\subsubsection*{Timelike entanglement} \label{sec:EW-L-adj-T} 
For two purely timelike adjacent subsystems $A=[(x,t_1),(x,t_2)]$ and $B=[(x,t_2),(x,t_3)]$ in the finite-sized $\TTbar$-deformed CFT$_2$, we may analytically continue our result for boosted spacelike intervals in \cref{EW-adj-Global}. Consequently the leading order expression for the bulk dual to the reflected entropy is given as follows
\begin{align}
	E_W(A:B) = &\frac{1}{4G_N}\left(\log\Bigg[\frac{2 L}{\pi\epsilon_c}\,\frac{\sin{\big(\frac{ \pi t_{12}}{L}\big)}\sinh{\big(\frac{\pi t_{23}}{L}\big)}}{\sinh({\frac{\pi t_{13}}{L} })}\Bigg]+i\pi\right)\notag\\
	&-\frac{\pi\chi_c^2 }{8 G_N L}\Bigg[t_{12} \cot \left(\frac{\pi t_{12}}{L}\right)+t_{23} \cot \left(\frac{\pi t_{23}}{L}\right)-t_{13} \cot \left(\frac{\pi t_{13}}{L}\right)\Bigg]\,,
\end{align}
where the first term signifies the contribution due to the undeformed CFT$_2$ and the rest of the terms proportional to $\chi_c^2$ denote the leading order corrections due to the $\TTbar$ with a small deformation parameter. Note that, above expression features a constant imaginary part reminiscent of a timelike curve in the corresponding bulk geometry. As earlier, upon utilizing \cref{rho-c-global} and the Brown-Henneaux formula, the above expression matches identically with half of the reflected entropy computed through conformal perturbation theory in \cref{SR-adj-L-T}.
\subsubsection{A single subsystem} \label{sec:EW-L-sing}
In this subsection, we consider a boosted spacelike subsystem $A=[(x_1,t_1),(x_2,t_2)]$ in the finite-sized CFT$_2$ with $\TTbar$ deformation. Since the spacelike direction $\phi$ is compactified, we do not encounter any pathology associated with an infinite branch cut. Hence the EWCS corresponding to the single subsystem under consideration reduces to the minimal surface homologous to the subsystem $A$. The length of the subsystem may then be obtain using standard AdS$_3$/CFT$_2$ techniques as follows \cite{Jiang:2023ffu}
\begin{align}
	E_W(A:A^c)&=\frac{1}{2G_N}\log\left[\frac{L}{2\pi\epsilon_c}\sqrt{2\cos\left(\frac{2\pi t_{12}}{L}\right)-2\cos\left(\frac{2\pi x_{12}}{L}\right)}\right]\notag\\
	&\qquad\qquad\qquad\qquad\qquad
	\qquad+\frac{\pi\chi_c^2}{4G_N L}\frac{t_{12}\sin\left(\frac{2\pi t_{12}}{L}\right)}{\cos\left(\frac{2\pi t_{12}}{L}\right)-\cos\left(\frac{2\pi x_{12}}{L}\right)}
\end{align}
Once again, the first term indicates the EWCS corresponding to the spacelike subsystem in the undeformed CFT$_2$ whereas the second term signifies the leading order correction due to the $\TTbar$ deformation which, upon utilizing the holographic dictionary in \cref{rho-c-global} and the Brown-Henneaux relation, may be seen to match with half of the corresponding reflected entropy obtained using conformal perturbation theory in \cref{SR-sing-L}.

\subsubsection*{Timelike entanglement} \label{sec:EW-L-sing-T}
Finally, we consider a timelike single subsystem $A=[(0,0),(x,t)]$ in the finite-sized $\TTbar$ deformed CFT$_2$. Since the subsystem is placed along the non-compact direction, in the reflected entropy computations in the dual field theory, we encounter an infinite branch cut. As earlier, we may circumvent this pathology by introducing two large auxiliary subsystems $B_1=[(-X,-T),(0,0)]$ and $B_2=[(x,t),(X,T)]$ sandwiching the single subsystem $A$. Considering the bulk entanglement wedge dual to the density matrix $\rho_{AB}$, we may obtain the upper bound of the EWCS between $A$ and $B$ utilizing \cref{EW-upper-bound}. In the following, we find it easier to employ Euclidean signature of the metric \cref{global-metric}, namely we make the Wick rotation to the Euclidean time: 
\begin{align*}
	t\to -i t_E~~,~~ T\to -i T_E\,.
\end{align*}
Now utilizing \cref{EW-upper-bound,EW-adj-Global} we obtain the upper bound as follows
\begin{align}
	\tilde{E}_W(A:B)=\frac{1}{4G_N}\log& \left[2\left(\frac{L}{\pi\epsilon_c}\right)^2\left(\cos \left(\frac{2 \pi  x}{L}\right)-\cosh \left(\frac{2 \pi  t_E }{L}\right)\right)\sqrt{\displaystyle\frac{\cos \left(\frac{2 \pi  (x-X)}{L}\right)-\cosh \left(\frac{2 \pi  (t_E -T_E)}{L}\right)}{\cos \left(\frac{2 \pi  (x+X)}{L}\right)-\cosh \left(\frac{2 \pi  (t_E +T_E)}{L}\right)}}\right]\notag\\
	&-\frac{\pi \chi_c^2}{8G_N L}\Bigg[\frac{2 t_E  \sinh \left(\frac{2 \pi  t_E }{L}\right)}{\cos \left(\frac{2 \pi  x}{L}\right)-\cosh \left(\frac{2 \pi  t_E }{L}\right)}+\frac{(t_E +T_E) \sinh \left(\frac{2 \pi  (t_E +T_E)}{L}\right)}{\cos \left(\frac{2 \pi  (x+X)}{L}\right)-\cosh \left(\frac{2 \pi  (t_E +T_E)}{L}\right)}\notag\\
	&\qquad\qquad\qquad\qquad\qquad\qquad\qquad~~~-\frac{(t_E -T_E) \sinh \left(\frac{2 \pi  (t_E -T_E)}{L}\right)}{\cos \left(\frac{2 \pi  (x-X)}{L}\right)-\cosh \left(\frac{2 \pi  (t_E -T_E)}{L}\right)}\Bigg]\label{blabla}
\end{align}
The upper bound to the EWCS for the single subsystem under consideration, may now be obtained by taking the bipartite limit $B\to A^c$ which may be achieved by stretching the auxiliary subsystems to (timelike) infinity, namely $T_E\to\infty$. Taking the appropriate limit of \cref{blabla} and analytically continuing to Lorentzian signature, we obtain
\begin{align}
	\tilde{E}_W(A:A^c)=\frac{1}{2G_N}&\left(\log\left[\frac{L}{2\pi\epsilon_c}\sqrt{\displaystyle 2\cos \left(\frac{2 \pi  x}{L}\right)-2\cos \left(\frac{2 \pi  t }{L}\right)}\right]+\log 2+i\frac{ \pi\,t}{L}\right)\notag\\
	&+\frac{\pi\chi_c^2}{4G_N L}t\left[2 i+\frac{\sin \left(\frac{2 \pi  t}{L}\right)}{\cos \left(\frac{2 \pi  t}{L}\right)-\cos \left(\frac{2 \pi  x}{L}\right)}\right]
\end{align}
The imaginary part of the (upper bound) to the EWCS may be rewritten in the following instructive form
\begin{align}
	\frac{\pi t}{2G_N L}\left(1+\frac{\chi_c^2}{2}\right)\,.
\end{align}
We notice the striking resemblance of the above expression to the $\TTbar$ corrected thermal entropy in \cref{Thermal-entropy}. However, the usual understanding in terms of wrapping around a horizon is missing. At this point, we boldly hypothesize that the presence of light cones may give rise to some kind of cosmological horizons impenetrable by the timelike segments of the corresponding HRT surfaces. For comparison with the field theoretic computations for the leading order corrections to the reflected entropy in \cref{SR-sing-L-T}, we reproduce the upper bound to the EWCS for a purely timelike subsystem $A=[(0,0),(0,t)]$, 
\begin{align}
	\tilde{E}_W(A:A^c)=\frac{1}{2G_N}&\left(\log\left[\frac{L}{\pi\epsilon_c}\sin \left(\frac{\pi  t }{L}\right)\right]+\log 2\right)\notag\\
	&-\frac{\pi\chi_c^2}{4G_N L}\frac{t}{\tan \left(\frac{\pi  t}{L}\right)}+i\frac{\pi t}{2G_N L}\left(1+\frac{\chi_c^2}{2}\right)\,.
\end{align}
Utilizing \cref{rho-c-global} and the Brown-Henneaux formula, the above expression conforms to the result obtained in \cref{SR-sing-L-T} through conformal perturbation theory till leading order (cf. \cref{footnote-single-BTZ}).

\section{Summary} \label{sec:summary}


In this article, we have performed a perturbative analysis to investigate the mixed state entanglement structure of $\TTbar$ deformed CFT$_2$s through the reflected entropy. In the regime of small deformation parameter $\mu \ll 1$, the partition function on the $nm$-sheeted replica manifold $\mathcal{M}_{nm}$ admits an expansion in $\mu$ with the leading term arising from the original (unperturbed) field theory. $\mathcal{O} (\mu)$ correction to the R\'enyi reflected entropy is then obtained by computing the expectation value of the $\TTbar$ operator in this manifold through conformal Ward identities. In the replica limit, we finally obtain the correction to the reflected entropy for disjoint, adjacent and a single boosted subsystem in thermal and finite-sized CFT$_2$s with $\TTbar$ deformation. Furthermore, we perform an analytic continuation of our result to timelike situations. Through this we investigate the nature of timelike entanglement for bipartite mixed states.

Note that, for the case of a single spacelike subsystem at a finite temperature and a single timelike subsystem in a finite-sized CFT, the replica structure is non-trivial and involves introduction of two auxiliary subsystems sandwiching the single subsystem in question. This is due to the presence of infinite branch cuts along the axis of the cylinder. Similar pathology has also been encountered for the reflected entropy in unperturbed CFT$_2$s earlier in the literature. Finally the reflected entropy is obtained for this configuration by implementing the bipartite limit where the auxiliary subsystems are pushed to the spatial and temporal infinity respectively. This pathology however does not arise for any other configuration as the subsystems are either finite or are aligned along the compactified direction of the cylinder.

We have also computed the bulk EWCS in dual AdS$_3$ geometries where the cut-off surface is pushed inside the bulk. The perturbative analysis is performed in terms of the location of this new cut-off surface. Specifically, we obtain the EWCS in cut-off BTZ black holes and cut-off global AdS$_3$ for disjoint, adjacent and a single boosted subsystem in the dual field theory. For the case of the cut-off BTZ black hole, rescaling of the spatial coordinate is required to appropriately map the location of the subsystems in the dual field theory as the coordinates are scaled along the holographic direction. However, for the cut-off global AdS$_3$ geometries, the temporal coordinate is scaled to match the CFT$_2$ defined on the new cut-off surface. With these, we observe perfect agreement of the bulk EWCS with the corresponding reflected entropy for all configurations considered. Furthermore, we also obtain the EWCS for timelike subsystems and observe proper matching with the corresponding field theory results. These serve as consistency checks for our computations. 

For the case of a single spacelike subsystem in a thermal CFT$_2$ deformed with the $\TTbar$ operator, we observe that the subtracted thermal contribution to the EWCS also receives $\mathcal{O}(\mu)$ correction which may be attributed to the decrease in the wrapping of the extremal curve around the black hole horizon as the holographic screen is pushed inside the bulk. A similar behaviour is also observed for a timelike subsystem in a finite-sized CFT$_2$ where a term, which looks suspiciously similar to the thermal entropy in finite temperature case, is subtracted from in the EWCS although no horizon is present in the dual bulk. An interpretation for this term is still unavailable to us but we suspect that the light cones may induce certain cosmological horizons which have this effect on the EWCS. We leave a concrete investigation of the origin of this behaviour for the future.

There are several open problems worth exploring in this direction. It will be interesting to study other mixed state entanglement and correlation measures such as the entanglement of purification, balance partial entanglement in these deformed theories. It would also be interesting to explore such irrelevant deformation in CFTs with unequal left-moving and right-moving central charges. One may also generalize our study to other excited states in CFT$_2$s such as through the addition of a $U (1)$ charge. Based on a recent article where the authors in \cite{Deng:2023pjs} have analysed the $\TTbar$ deformation of a boundary conformal field theory, it would be interesting to perform a similar analysis for reflected entropy and investigate the island formalism for these settings. We leave these interesting open issue for future consideration.

\section*{Acknowledgement}
VR would like to thank Lavish for several helpful discussions. Authors would like to thank the hospitality of IIT Mandi during the Student Talks on Trending Topics in Theory (ST4) 2023 conference where part of this work was completed.



\bibliographystyle{utphys}

\bibliography{reference}

\end{document}